\documentclass[fleqn,usenatbib]{mnras}
\pdfoutput=1

\usepackage{newtxtext,newtxmath}

\usepackage[T1]{fontenc}

\DeclareRobustCommand{\VAN}[3]{#2}
\let\VANthebibliography\thebibliography
\def\thebibliography{\DeclareRobustCommand{\VAN}[3]{##3}\VANthebibliography}

\usepackage{graphicx}	
\usepackage{amsmath}	

\title[Forming the First Water]{Constraining the Minimum Halo Mass that Supports Water Formation in a CCSN Remnant}

\author[C. T. D. Jessop]{
Christopher T. D. Jessop,$^{1}$\thanks{E-mail: christopher.jessop@port.ac.uk (KTS)}
\\
$^{1}$Institute of Cosmology and Gravitation, 1-8 Burnaby Rd, Portsmouth, PO1 3FX, UK\\
}

\date{Accepted XXX. Received YYY; in original form ZZZ}

\pubyear{2022}

\begin{document}
\label{firstpage}
\pagerange{\pageref{firstpage}--\pageref{lastpage}}
\maketitle

\begin{abstract}
We present a simulation probing the formation of water in the remnant of low-mass Population III supernovae in a cosmological minihalo, and provide a tentative lower mass limit on host minihaloes that can recollapse on a short enough timescale and efficiently mix metals at high densities. We start from cosmological initial conditions and end the simulation when the central density undergoes catastrophic recollapse, whereby the water abundance is reported. During the Population III stars lifetime, the minihalo (M$ = 5 \times 10^5$ M$_{\odot}$) becomes blown out, and consequently the faint supernova explosion (E$_{\mathrm{SN}} = 5\times 10^{50}$ ergs) is completely unconfined to the virial radius of the minihalo. At the end of the simulation there is no significant water formation anywhere throughout the remnant, and the central recollapsing region is inefficient at incorporating the first metals into itself, remaining at low metallicity. The majority of metals are ejected from the core via bipolar outflow into the void and reach a peak metallicity of $\mathrm{Z} \sim 10^{-6}\ \mathrm{Z}_{\sun}$ at very low densities. The mass of the minihalo is low enough such that the recollapse timescale is unreasonable for this configuration to be the primary avenue of water formation in the early universe. We also provide a comparison with a regular CCSN (E$_{\mathrm{SN}} = 10^{51}$ ergs) and find the same effect, but amplified. As such, we can suggest that the minimum minihalo mass required for a confined explosion, and therefore the possibility of water formation is at least $10^{6}$ M$_{\odot}$ and the chemo-thermal evolution of a supernova remnant is more sensitive to the mass of the host minihalo than the mass of the Population III star residing within it. 
\end{abstract} 

\begin{keywords}
hydrodynamics -- stars: Population III -- astrochemistry -- HII regions -- stars: low-mass
\end{keywords}



\section{Introduction}

The emergence of life in the universe is one of the most fundamental questions that arises to some degree within most areas and disciplines of science. A question of such complexity and depth cannot possibly be answered in a single paper, however, beginning to understand the cosmological conditions that are sufficient or even promote the inception of life is just one way we can begin to partly understand and answer this question. 

Water is an essential ingredient for life as we know it \citep{kasting2012find}, which is why an understanding of early-universe chemistry with an emphasis on numerous life-giving molecules, in particular water, is a crucial first step to take. There is currently very little in the way of hydrodynamical simulations or first principle calculations when it comes to modelling early universe water formation.  

Cosmological simulations on the collapse of primordial molecular clouds \citep{gnedin1997destruction} suggest that the first generation of stars, known as Population III stars (Pop III henceforth) may have had extremely large masses, anywhere within the range of $10\ \mathrm{M}_\odot$ to $>1000\ \mathrm{M}_\odot$. This first generation of stars are characterised by their extremely low metallicity, of at most $10^{-6}\  Z_\odot$ due to the availability of predominantly hydrogen and helium at their formation. Formation pathways for Pop III stars can be categorised into  high and low mass regimes. Firstly, metal-free molecular gas clouds cool very inefficiently and may result in Jeans-mass fragments on the order of $1000\ \mathrm{M}_\odot$ \citep[][]{barkana2001beginning,bromm2004first,bromm2009formation}. This leads to the possibility of either the formation of a free-growing dense core due to unopposed accretion from its host halo \citep[][]{abel2002formation,yoshida2006formation} or the formation of a secondary core \citep[][]{turk09}. Additionally, there exists a prediction for lower-mass Pop III star formation ($< 10\ \mathrm{M}_\odot$) under photodissociative feedback in systems where a minihalo forms at later redshifts, whereby the accretion rate of emerging stellar systems is seen to be a lot lower than stellar systems within minihaloes that have formed at earlier redshifts \citep{stacy2014first}.

Big Bang Nucleosynthesis allows only the production of the lightest elements, namely isotopes of hydrogen and helium, with trace amounts of lithium \citep{1995}. These elements underwent gravitational collapse and formed the first cosmological minihaloes, inside of which Pop III star formation occurred to mark the end of the cosmological dark ages. These stars are thought to be the first great nucleosyntethic engines in the universe, expelling vast quantities of heavier elements outwards and enriching both their own host minihalo, and other surrounding minihaloes. Depending on the mass of a Pop III star at the end of its lifetime, it is signified by a supernova that takes one of two types. Pop III stars in the mass range of $10 - 25\ \mathrm{M}_\odot$ explode as Type-II Core-Collapse Supernov\ae\ (CCSN) with a corresponding energy E$_{\textrm{SN}} = 0.5 \times 10^{51}$ ergs and leave behind a compact remnant. At the higher end of the Pop III mass spectrum, stars within the mass range of $140 - 260\ \mathrm{M}_\odot$ end their lives as Pair-Instability Supernov\ae\ (PISN) releasing an order of magnitude greater energy, and is completely disrupted leaving behind no remnant \citep{heger2003massive}. The most massive stars at the top of this range, i.e. $250\ \mathrm{M}_\odot$ and above, undergo complete photodisintegration. This reaction absorbs excess energy before runaway collapse leads to a hypernova, before collapsing as a black hole \citep{2001PISN}. Both CCSN and PISN forge the heavier elements crucial for complex molecule formation in their core, such as oxygen.
Chemical enrichment can be categorised into 3 distinct methods depending on the mechanisms at play. Primarily, metals can be incorporated into massive haloes via a hierarchical structure formation process. \citet{get07} explore this scenario where they explode a $200\ \mathrm{M}_\odot$ PISN, which expels an interior, metal enriched, bubble that expands adiabatically into the void and inter-galactic medium (IGM) through cavities created by the supernova shock. It is concluded that a dark matter halo with a virialised mass $> 10^{8}\ \mathrm{M}_\odot$ is required to recollect the shocked gas and allow mixing to occur. The second method of metal enrichment of a halo can be referred to as the IE mechanism. This mechanism refers to the way enriched material falls back into the host halo after a period of recovery, and is applicable to describe the sequence of metal enrichment only when the supernova progenitor is on the order of tens of solar masses. The reason for this is to ensure that the host halo has not been blown out as described previously, and so that the HII (Singly ionised Hydrogen) region is relatively confined to well within the virial radius. It is reported that a moderately massed progenitor, and by extension, a moderate energy supernova, initially is only able to photoionise a small region within its host halo, i.e. a trapped HII region. Following the instability created due to the supernova explosion, ejection material begins to fall back to the central region, reaching a steady accretion rate after approximately $5$ Myr, and less than half of the ejecta is able to escape the virial radius. \citep{ritt12, rit16}.
A relatively new concept for metal enrichment was explored to explain the observations of the most metal-poor stars in a study that looked to describe the chemical abundance patterns of Carbon Enhanced Metal Poor (CEMP) stars, in particular, one with an Iron-to-Hydrogen ratio, [Fe/H] $= -5.54$ \citep{nor13}. Minihaloes that lie external to a Pop III stars host minihalo can undergo external enrichment (EE), whereby enriched supernova material escapes the host minihalo and impacts these nearby minihaloes, potentially paving a way for the next generation of Population II (Pop II) star formation as long as these externally located minihaloes have not yet formed stars of their own. The pioneering study to look specifically at the EE mechanism, concludes that turbulence from the virialisation of the impacted minihalo is able to incorporate enriched material into a $0.01$ pc radius to a uniform metallicty Z$\ \sim 2\times10^{-5}\ \mathrm{Z}_\odot$ \citep{smith2015first}.

The possibility of early-universe water formation has not been vigorously tested until relatively recently. \citet{2015Bialy} demonstrate that in the low-metallicity environments ($Z = 10^{-3}\ Z_\odot$) typical of the metal enrichment epoch, significant quantities of water vapour could have been present, using idealised one-dimensional calculations in a system of partially shielded gas. Water formation in the present day is dominated by reactions that occur on the surface of dust grains, for example through the hydrogenation of oxygen \citep[][]{2014Dishoeck}. Due to the lack of metals that constitute dust grains in the early universe, water formation on dust is not a reasonable avenue to explore at these higher redshifts, and we therefore must look at alternative formation pathways. At temperatures $\geq 300$ K, water is able to form via gas-phase neutral-neutral reactions. Hydroxyl (OH) is initially formed when oxygen reacts with H$_2$, which in turn reacts with H$_2$ forming water. 

\begin{equation}
    \textrm{O} + \textrm{H}_2 \rightarrow \textrm{OH} + \textrm{H}
\end{equation}

\begin{equation}
    \textrm{OH} + \textrm{H}_2 \rightarrow \textrm{H}_2\textrm{O} + \textrm{H}
\end{equation}

The conditions for this formation pathway are thought to be typical of shocks, where most of the oxygen can be driven into water where the gas has heated to higher temperatures \citep[][]{1983ApJ...264..485D}. To summarise what is required within a region for water formation to occur in the primordial universe, there needs to be a region of warm, dense gas at a relatively high metallicity (there needs to be sufficiently enough oxygen)

\section{Method}

We run our simulations using the  \textsc{Enzo} v2.5 simulation code \citep{o2005introducing,bryan2014enzo}.  \textsc{Enzo} is an open-source, Eulerian, adaptive mesh refinement (AMR) code that solves N-body dark matter dynamics. The equations of hydrodynamics are solved using a second-order piecewise parabolic method (PPM), fully detailed in \citet{colella1984piecewise}. Conventionally,  \textsc{Enzo} uses a fixed value for the adiabatic index $\gamma = 5/3$ if the selected hydro method is PPM. However, at very low metallicities $Z_{\odot} < 10^{-3}$ and at densities n$_{\textrm{H}} \geq 10^8$ cm$^{-3}$, the adiabatic index approaches $\gamma = 7/5$ as the gas becomes fully molecular. Therefore, the hydrodynamics solver is extended to consider a variable adiabatic index.  

\subsection{Chemistry - Heating and Cooling}
 \textsc{Enzo} natively is limited to the primordial 12-species model: e$^-$, H, H$^+$, He, He$^+$, He$^{++}$, H$_2$, H$_2^{+}$, H$^{-}$, D, D$^+$, and HD. These species can be solved either within  \textsc{Enzo's} internal chemistry solver or with the external chemistry and radiative cooling package  \textsc{Grackle}, which can be used to advect and evolve non-equilibrium primordial chemistry \citep{2017MNRASGrack.466.2217S}. Metal cooling rates are interpolated from tables using the  \textsc{CLOUDY} \citep{smith2008metal} cooling photoionisation code, which are four-dimensional tables that interpolate over density, metallicity, electron fraction, and temperature. Any further chemistry, such as molecular chemistry will require a significantly expanded chemical network. A modified  \textsc{Grackle} scheme has been implemented \citep{10.1093/mnras/sty2984} and extended to include 49 reactions for the primordial species listed above, in addition to HeH$^+$, D$^{-}$, and HD$^+$ to create a 15-species primordial model. This model includes various processes such as collisional ionisation and recombination of H$_2$, in addition to its formation and destruction pathways from three-body reactions. Primordial heating and cooling processes are included, such as gas heating. In the temperature regime T$\ > 8000$ K, inverse Compton cooling, brehmsstrahlung, H and He species transitional line cooling, and ionisation/recombination processes are considered. For temperatures under $10000$ K, the contribution of molecular cooling from H$_2$ and HD are calculated. The hydrogen mass fraction remains constant at $X_{H} = 0.76$, whereas the helium fraction varies in the region surrounding the SN ejecta where it becomes enhanced. 
The effect of metals when first introduced from Pop III stars have a profound impact on the evolution of a halo \citep{bromm2003formation}. Therefore to follow these effects in detail, the extra species and reactions are taken from \citet{2005Omukai} and added to the extended version of  \textsc{Grackle}. The additional species being considered are: C, C$^+$, O, O$^{+}$, CH, CH$_{2}^{+}$, CO, OH, H$_2$O, O$_2$, CO$^{+}$, O$_2^{+}$, OH$^{+}$, H$_2$O$^{+}$, H$_3$O$^{+}$, CO$_2$, Si, SiO, and SiO$_2$ for a total of 40 non-primordial reactions. The cooling due to fine-structure level transitions for C, C$^+$, and O are included, in addition to the rotational transitions of H$_2$O, OH, and CO by interpolating over pre-computed tables. 

\subsection{Dust Treatment}
Similarly to the metal species, the chemical network has been extended to eight species of dust: both metallic silicon (Si) and metallic iron (Fe), forstertite (Mg$_2$SiO$_4$), magnesia (MgO), enstatite (MgSiO$_3$), silica (SiO$_2$), amorphous carbon (C), and troilite (FeS). \cite{smith2015first} demonstrate the significance that dust has on the thermal properties of a gas cloud. Dust can very effectively enhance gas cooling which in turn initiates fragmentation. The thermal effects of dust on a gas cloud can be discretised in four ways: 
\begin{enumerate}[i]
    \label{table:dust_list}
    \item Formation of H$_2$ on dust grains (crucial to consider for water formation)
    \item Gas cooling via thermal emission from dust grains
    \item Accretion of metals in the gas-phase onto dust grain surfaces
    \item Creates a continuum opacity
\end{enumerate}
A detailed description of the formulation for each rate in \ref{table:dust_list} is provided in \cite{10.1093/mnras/stu2298}. Each grain species has its rates tabulated in its own table, and these rates are interpolated from the gas temperature T, density $\rho$, density of a grain species $i (\rho_i)$, and the density of the species corresponding key element $X (\rho_X)$ at each timestep for a fluid cell. The continuum opacity is approximated as $\tau_{\textrm{cont}} = (\kappa_{\textrm{p}} \rho + \sum_i \kappa_i \rho_i) l_{\textrm{jeans}}$, where $\kappa_{\textrm{p}}$ is the Planck mean opacity of primordial gas, $\kappa_i$ is the mean opacity of a given grain species $i$, $\rho_i$ is the mass density of again, a grain species $i$, and the $l_{\textrm{jeans}}$ term is the Jeans length, which is used as a shielding length. A diffusion approximation that effects the continuum cooling rate processes such as dust thermal emission and collisionally induced excitation are reduced by a factor $\beta_{\textrm{cont}} = \textrm{min}\{1, \tau^{-2}_{\textrm{cont}}\}$.

\subsection{Simulation Setup}
We initialise the primary isolated minihalo runs using the  \textsc{MUSIC} initial conditions generator, an algorithm that generates multi-scale initial conditions with multiple levels of refinement for cosmological simulations \citep{hahn2011multi}. We initialise a $0.25\ $ Mpc/h comoving box using the Planck 2016 cosmological parameters $[\Omega_{m} = 0.3089, \Omega_{\lambda} = 0.6911, \Omega_{b} = 0.04860, H_{0} = 67.74, \sigma_{8} = 0.8159, n_{s} = 0.9667]$ \citep{ade2016planck}. For our exploratory run we use a resolution of $256^3$ corresponding to a dark matter mass resolution of $67.27\ \mathrm{M}_\odot$, with a refinement level $l = 1$ from $z = 200$ to $z = 15$ to find a minihalo with a total mass of $10^{6}\ \mathrm{M}_\odot$ using the  \textsc{ROCKSTAR} halo finder \citep{behroozi2012rockstar}. Simulations are then set up using the same random seed generators and initial conditions as the exploratory run, centred around the $10^{6}\ \mathrm{M}_\odot$ minihalo, and placing a nested grid with a resolution of 1024 zones. This target minihalo at SF will be approximately $10^{5}\ \mathrm{M}_\odot$ and be classed as a low-mass minihalo. The simulation is started from $z = 200$ with the added chemistry, radiative feedback and star formation mechanisms described above, with data being output at every time step, initially in code units of 5.0. Adaptive mesh refinement of cells occurs by a factor of 2 when the following criteria are met:
\begin{enumerate}
    \item Gas mass is $>4\times\bar{M}_\textrm{{baryon}}\times2^{-l\times0.2}$, where $\bar{M}_\textrm{{baryon}}$ is the average gas mass in a cell and l is the refinement level
    \item The dark matter in a cell is $>4\times{M_\textrm{{initial}}}$, where ${M_\textrm{{initial}}}$ is the initial mass of a cell
    \item Local Jeans length is $<16$ cells wide
\end{enumerate}
The reason for employing such criteria is as follows. Firstly, we have a negative exponent in the baryon density function allowing a greater rate of refinement with increasing levels of adaptive mesh refinement, ensuring super-Lagrangian behaviour \citep{o2007population}. The local Jeans length cell width constraint ensures that the Truelove criterion, which states that Jeans length has to be resolved by a minimum of 4 cells on each axis, is satisfied \citep{true97}. This ensures that there is no occurrence of artificial fragmentation. A refinement level $l = 15$ is set which is reached before the formation of the first Pop III star in our simulation, and eventually gives the required resolution to adequately show metal mixing and halo enrichment from SNe and consequent water formation. Analysis of all data and all images created from the raw data is done through the  \textsc{YT project}, a community developed, open source astrophysical analysis and visualization toolkit developed in Python \citep{turk2010yt}.

\subsection{Star Formation}
We model star formation and the resulting feedback of Pop III stars using the \cite{cen1992galaxy} algorithm extension \citep{wise2008resolving} built into  \textsc{Enzo}. This extension inserts a star particle into a grid cell when the following criteria are met:

\begin{enumerate}
    \item An overdensity $\geq 10^6$ cm$^{-3}$
    \item A converging gas flow/velocity field, i.e. $(\nabla\cdot\textbf{v}_{\textrm{gas}}<0)$
    \item A molecular hydrogen (H$_2$) fraction $>5\times10^{-4}$
    \item The metallicity is less than some critical metallicity value $(Z < 6 \times 10^{-8}$)
    \item The cooling time is less than the dynamical time (T$_{\textrm{cool}} < t_{\textrm{dyn}}$)
\end{enumerate}

These conditions are thought to be typical of metal-free clouds undergoing collapse approximately 10 Myr before the birth of a Pop III star \citep{abel2002formation}. The criteria regarding the molecular hydrogen fraction are required to limit star formation to molecular hydrogen clouds with large rates of H$_{2}$ formation relative to H$_{2}$ radiative dissociation. Once the conditions are met, a star particle will then form 10 Myr later, whereby the mass of the star is uniformly removed from a sphere containing twice the stellar mass, centred on the star particle. The Pop III star mass is sampled from the IMF
\begin{equation}
    \label{eq:IMF}
    f(\textrm{log} M_{\textrm{Pop III}}) = M^{-1} \textrm{exp} \left[- \left(\frac{M_{\textrm{char}}}{M_{\textrm{Pop III}}}\right)^{1.6}\right],
\end{equation}
where M$_{\textrm{char}} = 20\ M_{\odot}$, which is the characteristic mass of Pop III stars. The radiation field evolution from point sources, as in the case of Pop III stars, can be accurately calculated using adaptive ray tracing \citep{abel2007h}, utilising the MORAY radiation field solver \citep{wise2011enzo+}. Each individual star particle is treated as a point source of ionising radiation, where the photons in each ray are equal to one another. Furthermore, the sum of the photons in the initial rays are equal to the stellar luminosity. On incidence with a higher resolution AMR grid, the primary ray is split into 4 daughter rays if and when the associated solid angle $\theta > 20\%$ of the area of the cell. At the end of the stars lifetime, T$_{\textrm{life}}$, the supernova energy, E$_{\textrm{SN}} = 0.5 \times 10^{51}$ ergs is uniformly injected into a sphere with a radius of 10 pc. The metals corresponding to that of a progenitor are likewise assumed to be uniformly mixed within the ejecta. At a radius of 7.5 pc, a contact discontinuity (CD) is placed which contains all metals from the explosion. Only when the shock undergoes Rayleigh-Taylor (RT) instabilities can the metals break through the CD.

\subsection{Supernov\ae\ Yields}

\begin{figure}
    \centering
    \includegraphics[width=1.0\linewidth]{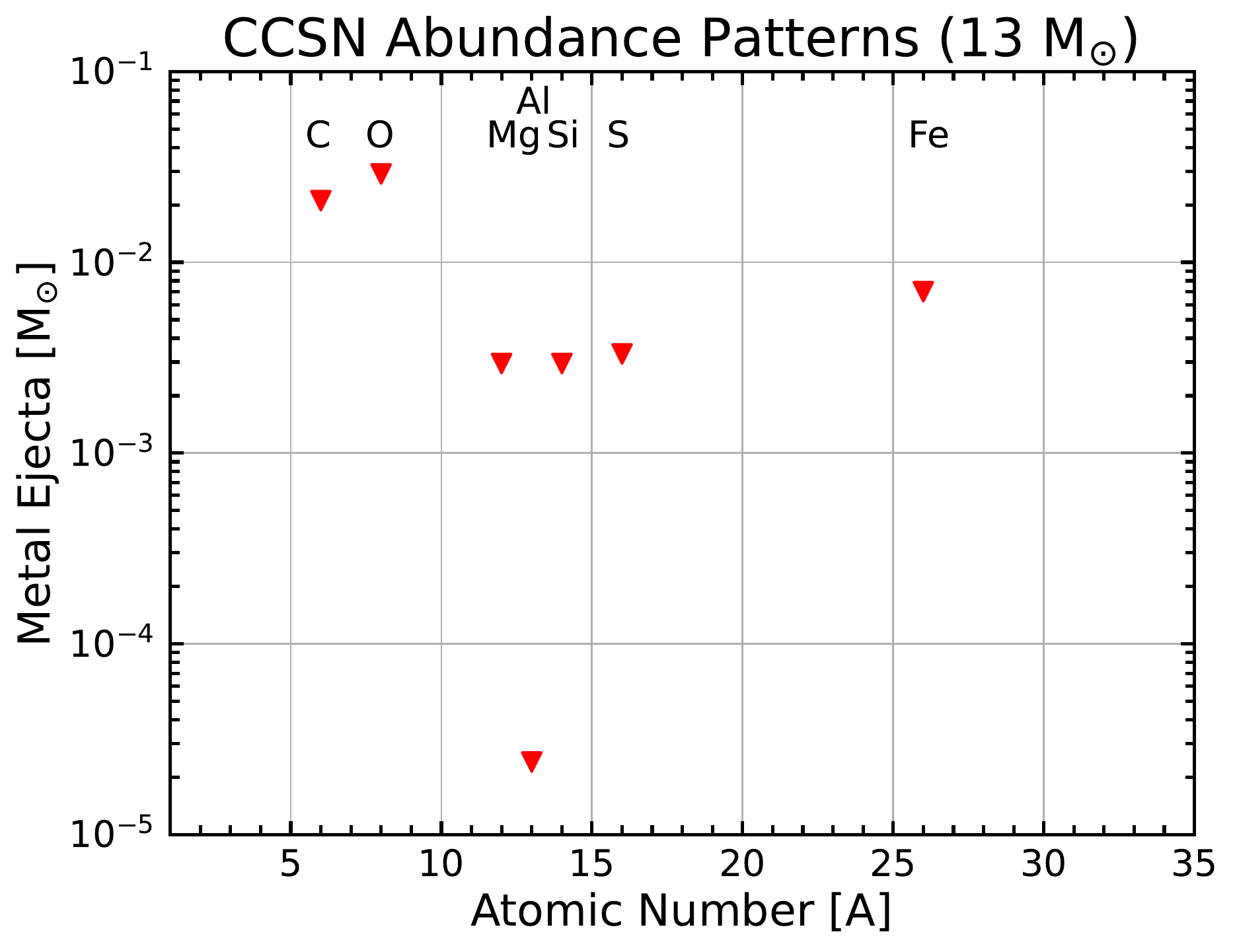}
    \caption{Metal ejecta in solar masses of the major elements for the $13$ M$_{\odot}$ faint-SN model used in this simulation.}
    \label{fig:CCSN_abundance}
\end{figure}

\label{sec:SN Yields}
The metal and dust abundances, as well as the dust size distribution are derived from \cite{Umeda&Nomoto2002, nozawa2007evolution} for our Pop III progenitor. The model considers SN yields for the following core-collapse masses: 13, 20, 25, and 30 M$_{\odot}$. In addition to the pair-instability masses: 170 and 200 M$_{\odot}$. For our purposes, only the $13\ \mathrm{M}_{\odot}$ model is selected and explodes with energy E$_{\textrm{SN}} = 0.5 \times 10^{51}$ ergs and outputs M$_{\mathrm{met}} = 0.097$ M$_{\odot}$ of metals. This represents a `subdued' SNe, where fallback onto the core is significant and dampens the effect of the explosion. Pop III metal abundance differ somewhat to the present-time ISM, specifically they show an alpha-species enhancement relative to that of the solar abundance ratio of $[\alpha / \textrm{Fe}] \simeq 0.4$. Note that in the case of a $13\ \mathrm{M}_{\odot}$\ CCSN, magnesium $([\textrm{Mg}/ \textrm{Fe}] = -0.23)$ and oxygen $([\textrm{O}/ \textrm{Fe}] = -0.15)$ abundances are relatively depleted with respect to the solar ratio, whilst sulphur $([\textrm{S}/ \textrm{Fe}] = 0.17)$ and silicon ($[\textrm{Si}/ \textrm{Fe}] = 0.11)$ are enhanced. 

\section{Results}

\begin{figure*}
    \centering
    \includegraphics[trim={0 4.1cm 0 0},scale=0.4]{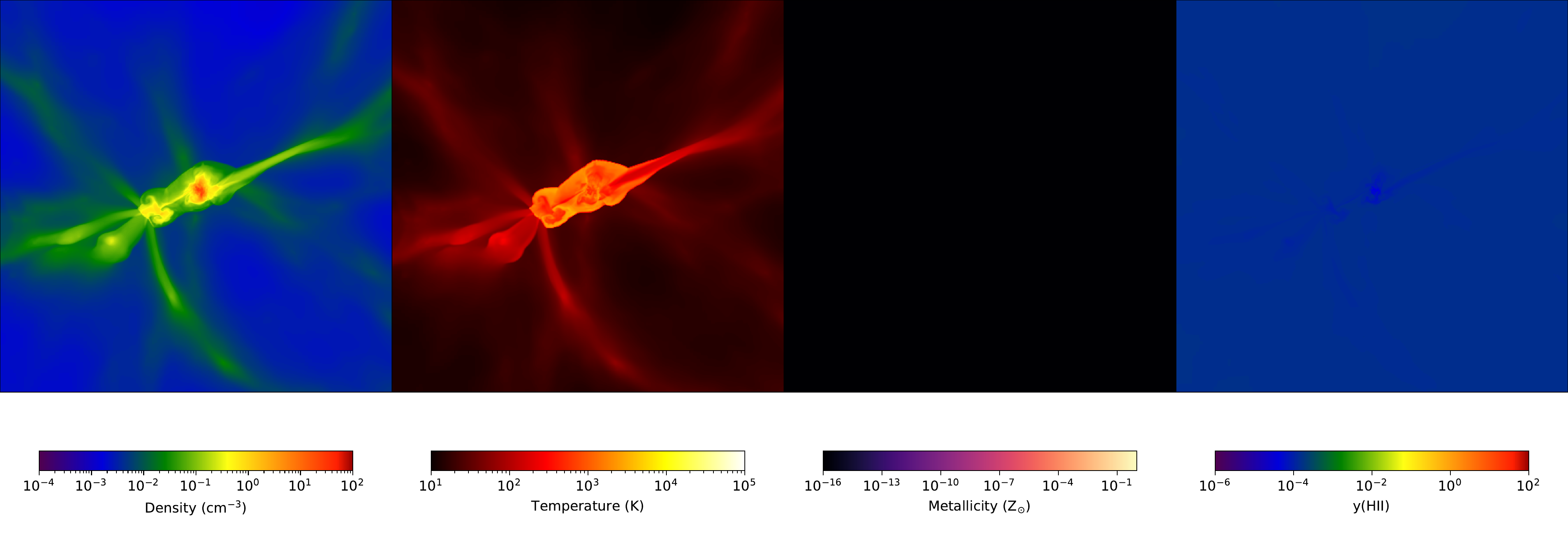}\label{subfig:multi_a}
    \includegraphics[trim={0 4.1cm 0 0},scale=0.4]{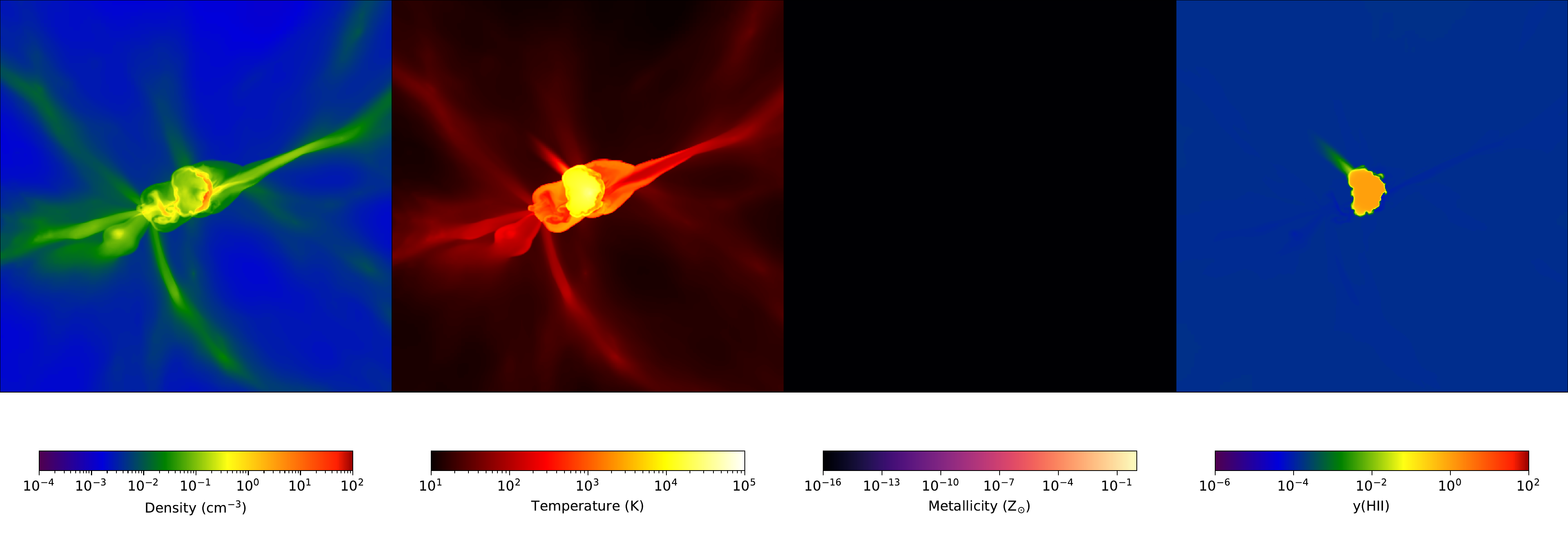}\label{subfig:multi_b}
    \includegraphics[trim={0 4.1cm 0 0},scale=0.4]{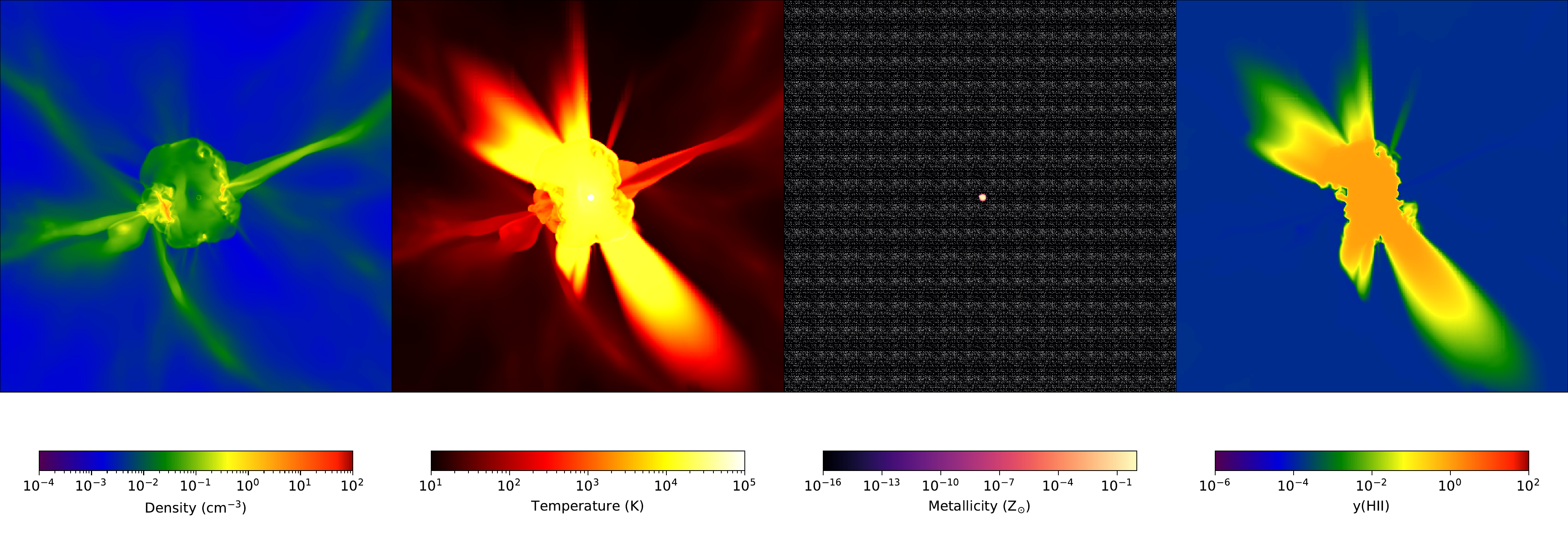}\label{subfig:multi_c}
    \includegraphics[trim={0 4.1cm 0 0},scale=0.4]{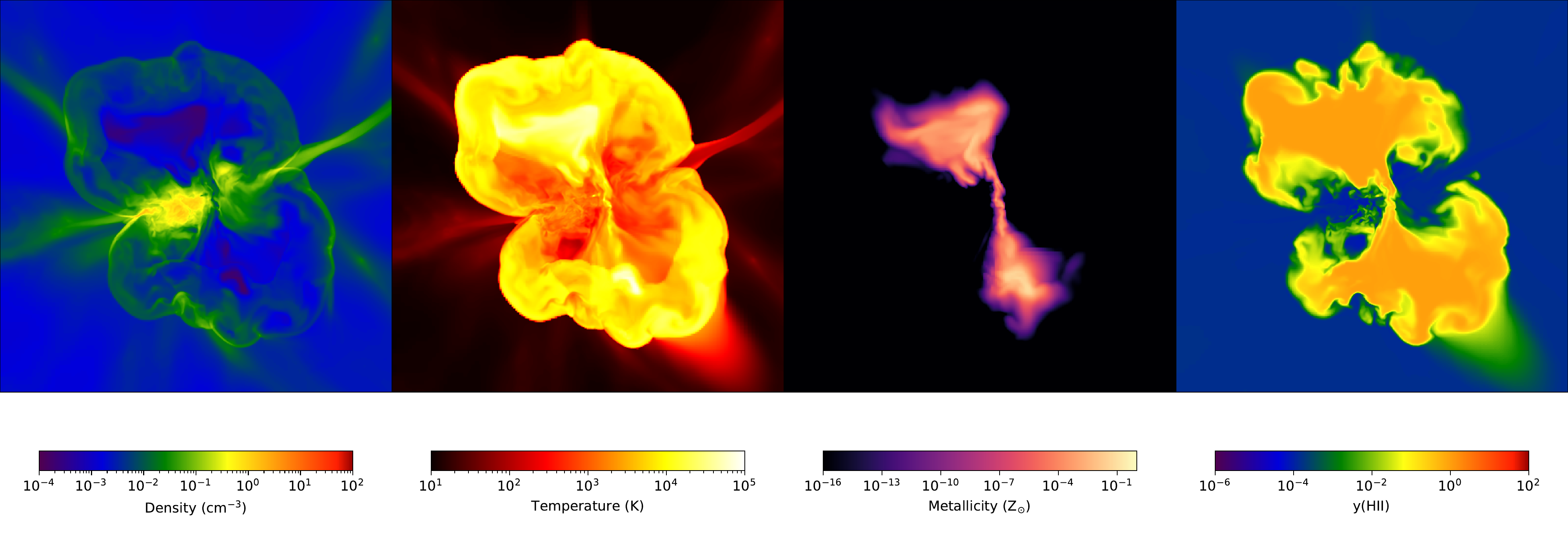}\label{subfig:multi_d}
    \includegraphics[scale=0.4]{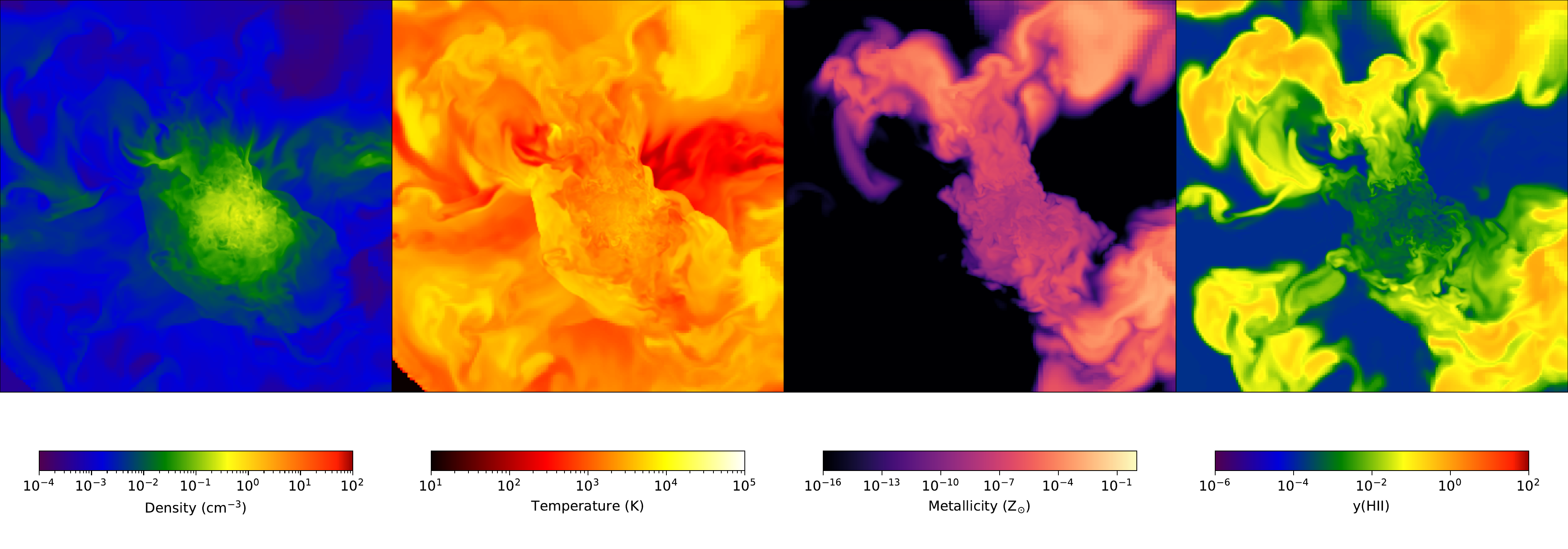}
    \caption{From left to right, slices of density, temperature, metallicity, and HII abundance centred on the coordinates of SF and SNe with a side length of 2 kpc comoving. From top to bottom, immediately prior to SF ($z = 19.10$), SF ($z = 18.53$), 1 Myr post-SNe ($z = 18.17$), 18 Myr post-SNe ($z = 17.41)$, and 112 Myr post-SNe ($z = 13.34$).}
    \label{fig:multi_slice}
\end{figure*}

\begin{figure*}
    \centering
    \includegraphics[scale=0.3]{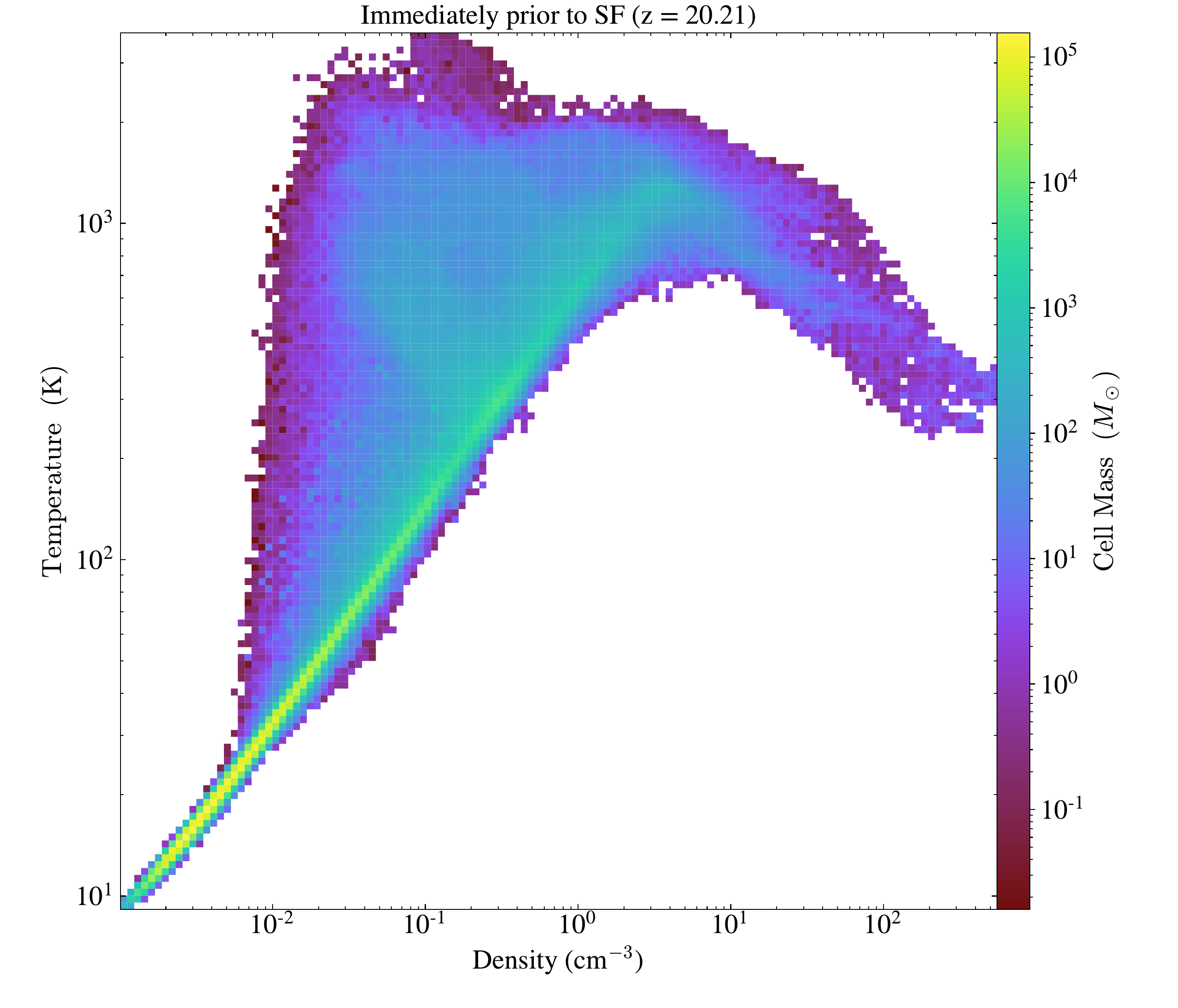}
    \includegraphics[scale=0.3]{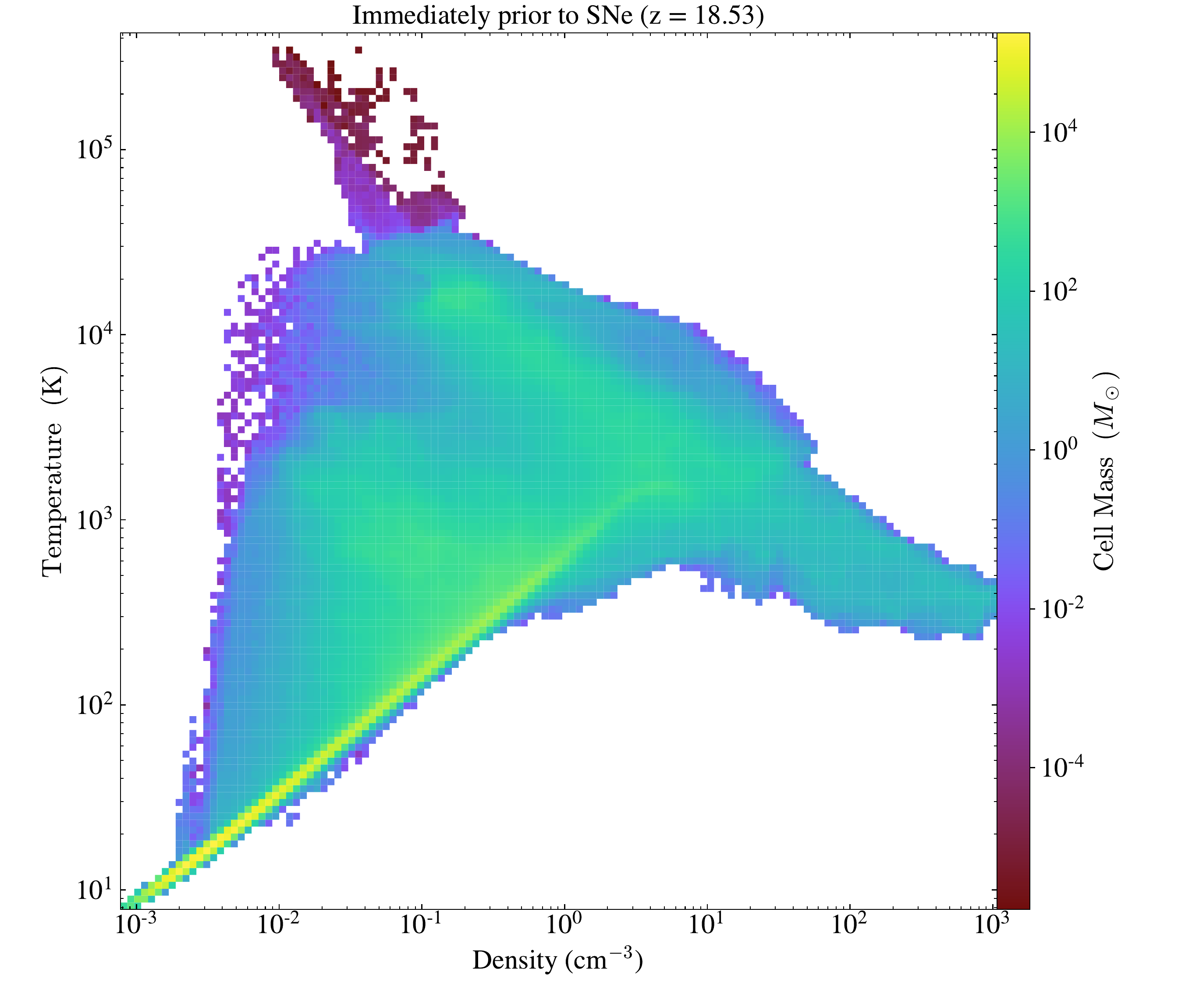}
    \includegraphics[scale=0.3]{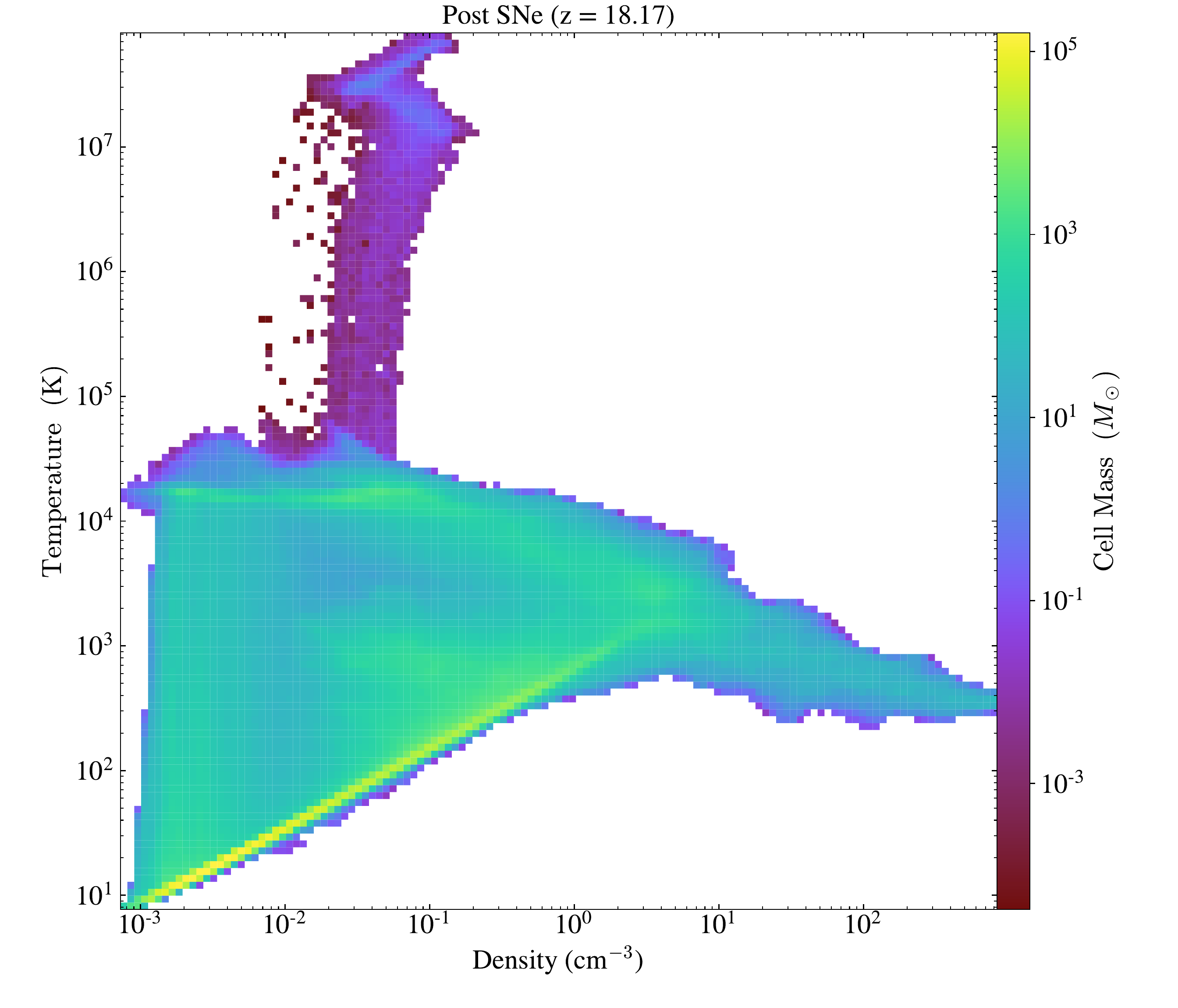}
    \includegraphics[scale=0.3]{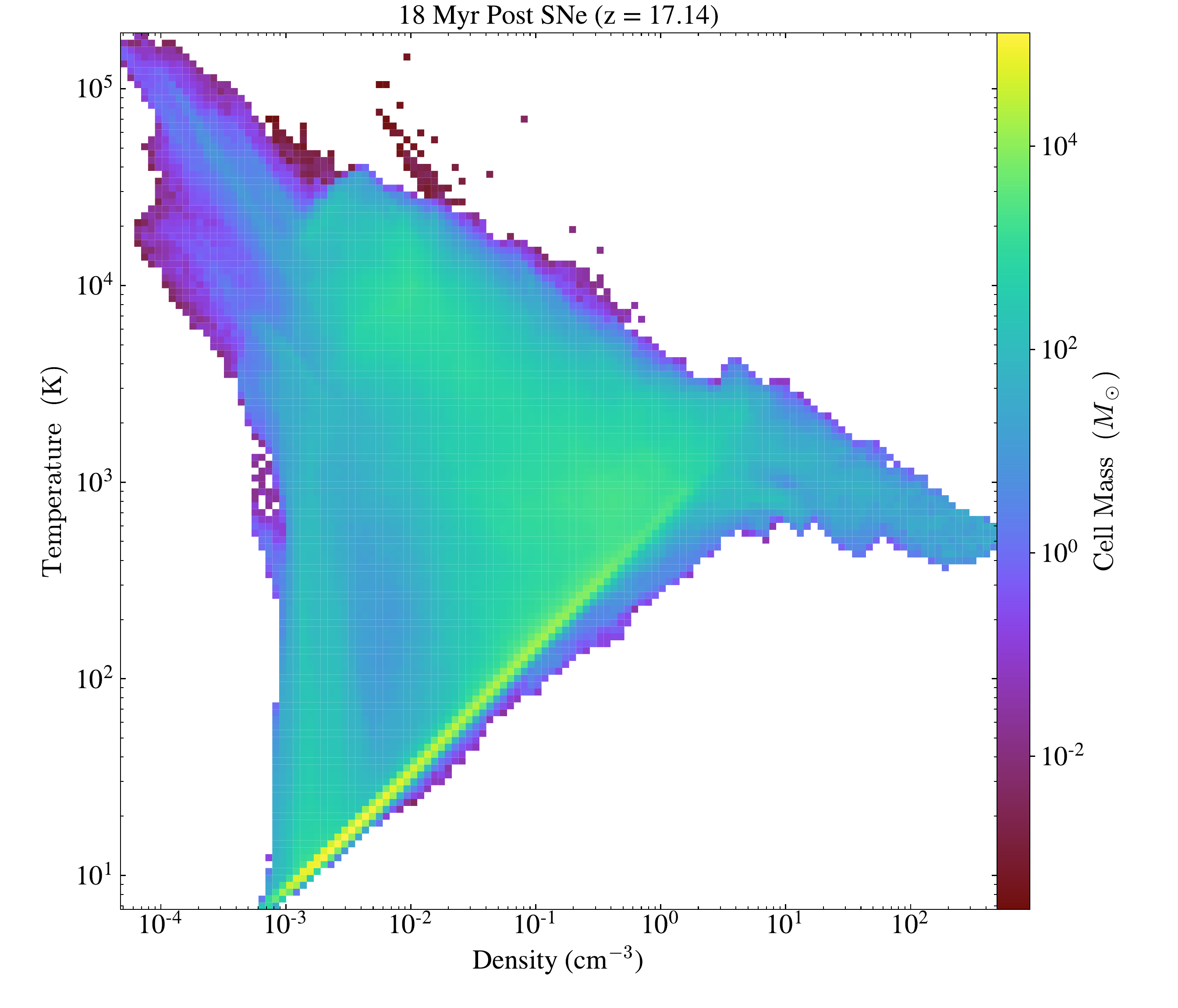}
    \includegraphics[scale=0.3]{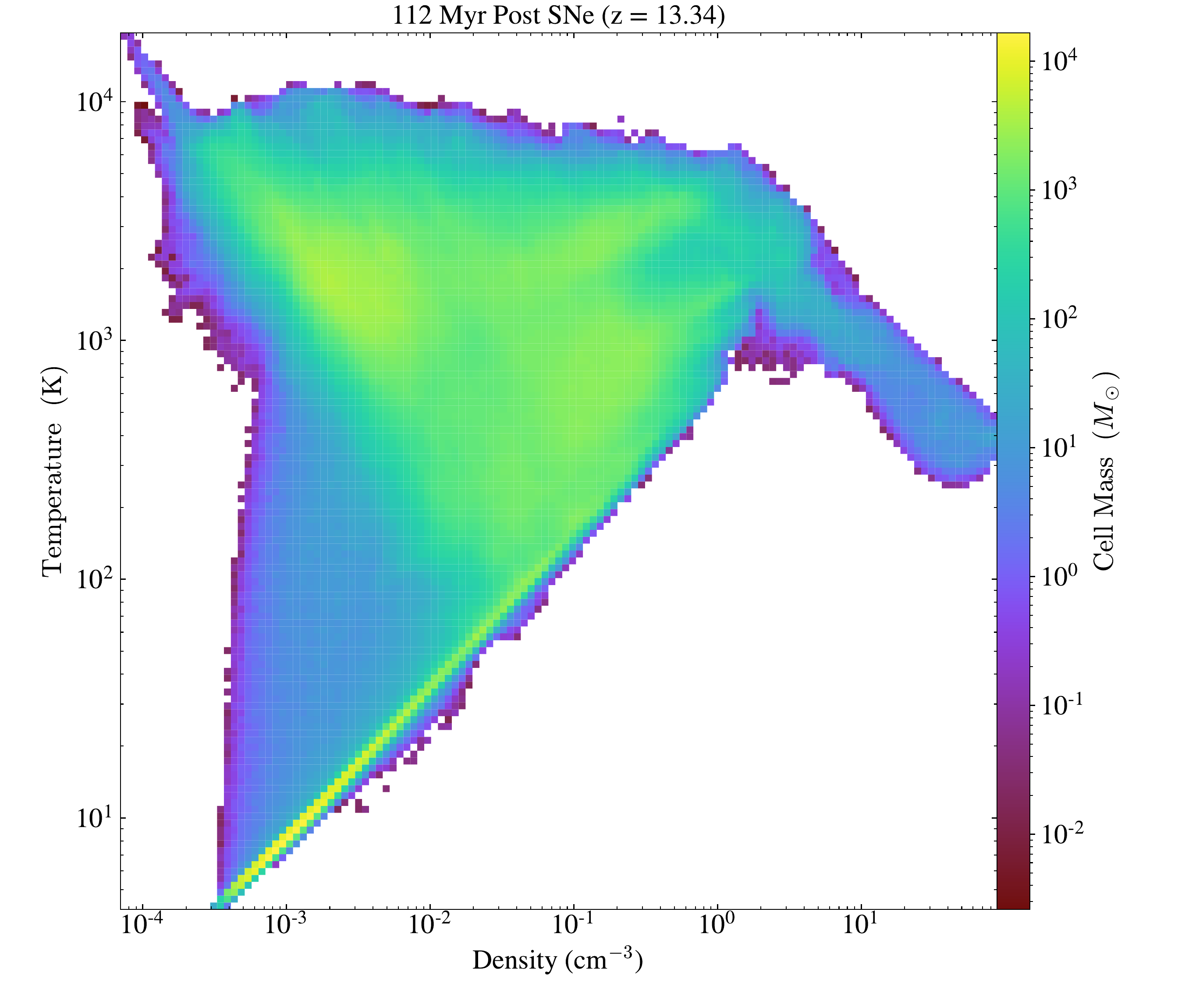}
    \includegraphics[scale=0.3]{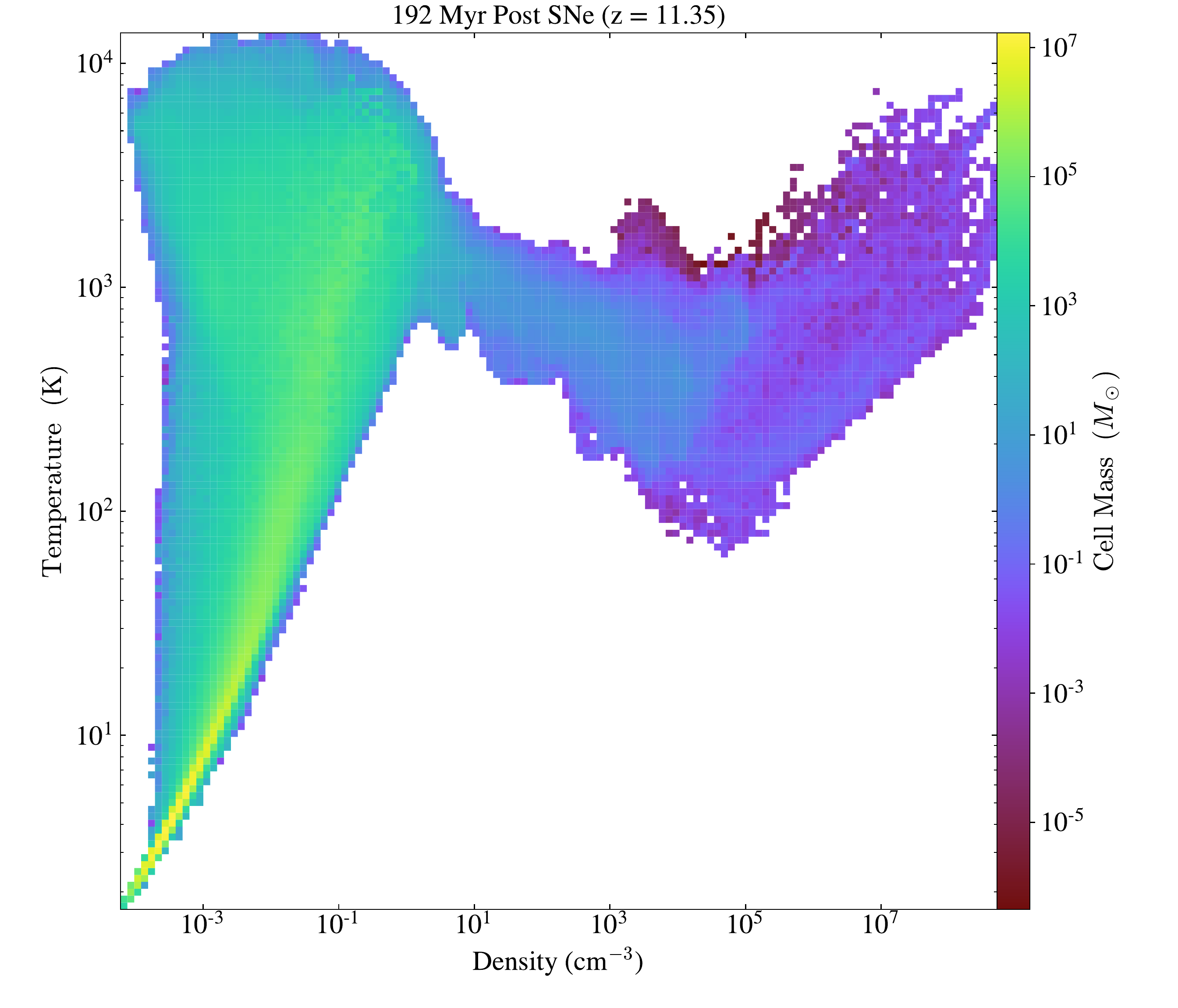}
    \caption{Two-dimensional phase plots displaying temperature and density at the same epochs as Fig. \ref{fig:multi_slice}.}
    \label{fig:phase_plots}
\end{figure*}

The entire sequence of minihalo collapse, Pop III star formation and HII region generation, supernova explosion feedback, and finally re-collapse are followed in detail. Fig. \ref{fig:multi_slice} display slice plots of density, temperature, metallicity, and HII abundance centred on the coordinates of SF. Each row of the figure corresponds to a significant period in the evolution of the system. Firstly, the state of the minihalo immediately prior to SF, after virialisation and gas collapse via HII cooling has created conditions congruent to SF (Fig. \ref{fig:phase_plots} Top left), shortly followed by SF and evolution itself (Fig. \ref{fig:phase_plots} Top right). The creation of the HII region resultant of the Pop III star is followed, which includes breakout regions that extend past 1 kpc, in addition to a D-type ionisation front (I-front). After the lifetime of the star T$_{\textrm{life}}$, has elapsed the CCSN occurs, whereby the first source of metals and dust are produced and expelled into the immediate environment (Fig. \ref{fig:phase_plots} Middle left).   

\subsection{Star Formation and Feedback}

The top row of Fig. \ref{fig:multi_slice} displays the state of the minihalo immediately prior to SF. Gas has accreted onto the minihalo and collapsed to densities around n$_{\textrm{H}} = 1$ cm$^{-3}$, at which point the cooling process is dominated by H$_2$ through rotational and vibrational transitions (ro-vibrational lines) until the temperature decreases to 200 K, and the density increases to n$_{\textrm{H}} \sim 10^{3}$ cm$^{-3}$. At this point, the ro-vibrational levels of H$_2$ are fully populated at their equilibrium levels and an apparent hydrostatic equilibrium is reached. Additionally, the cooling rate becomes independent of density. Early simulations for the collapse of primordial minihaloes \citep{gnedin1997destruction} suggest that neutral, metal-free pristine gas cools extremely inefficiently, and therefore result in Jeans-mass scale fragments that could reach $> 1000\mathrm{M}_\odot$ \citep{omukai2000protostellar,barkana2001beginning,bromm2004first,bromm2009formation}. In this scenario, the fragment results in a Pop III star formed with a mass $M_{\textrm{PopIII}} = 13\ \mathrm{M}_{\odot}$ and corresponding lifetime T$_{\textrm{life}} = 12$ Myr that forms at redshift $z = 19.02$. The Pop III star mass is sampled from the IMF given by eq. \ref{eq:IMF} in this instance. Immediately prior to SF, the mass of the minihalo is $M_{\mathrm{halo}} = 4.8 \times 10^{5}$ M$_{\odot}$. During its main-sequence lifetime (MSL), the star particle isotropically emits ionising and H$_2$ dissociating Lyman-Werner photons (LW) with rates of $Q(\mathrm{H}) = 1.09 \times 10^{48}$ s$^{-1}$ and $Q(\mathrm{LW}) = 1.54 \times 10^{48}$ s$^{-1}$ respectively \citep{schaerer2002properties}. The ionisation rates are given by
\begin{equation}
    Q(\mathrm{H}) = 10^{43.61 + 4.9x - 0.83x_{2}},
\end{equation}
\begin{equation}
    Q(\mathrm{H0}) = 10^{42.51 + 5.69x - 1.01x_{2}},
\end{equation}
\begin{equation}
    Q(\mathrm{He}) = 10^{26.71 + 18.14x - 3.58x_{2}},
\end{equation}
\begin{equation}
    Q(\mathrm{LW}) = 10^{44.03 + 4.59x - 0.77x_{2}},
\end{equation}
where $x = \mathrm{log}_{10}(\mathrm{M}_{\mathrm{star}})$ and $x_2 = x^2$.
The effects of the progenitor star are visible in both Fig. \ref{fig:multi_slice} and \ref{fig:phase_plots}. Specifically, Fig. \ref{fig:phase_plots} shows the formation of the low density, diffuse region surrounding the star, where the temperature increases to T$ > 10^{5}$ K, at a minimum density of n$_{H} \sim 10^{-3}$ cm$^{-1}$. Within this region, hydrogen has become fully ionised, and this becomes clearly visible in Fig. \ref{fig:multi_slice}, as it presents itself as the `butterfly' structure. The compact D-type I-front is visible within the density slice plot extending to approximately $100$ pc in each direction although the structure remains inhomogeneous, whereas it remains somewhat hidden by the breakout regions in the temperature slice plot. Along the `principal' axis (main horizontal structure when looking at Fig. \ref{fig:multi_slice}), the density remains relatively high (n$_{H} \sim 10^{-1}$ cm$^{-1}$) throughout as the edges of the minihalo meets the filamentary structure, compared to the regions above and below the minihalo that borders the void (n$_{H} \leq 10^{-2}$ cm$^{-1}$). As the main structure of the minihalo meets the voids, Lyman continuum leakage (LyC leakage) extends the HII region, as LyC photons are able to escape through holes of low column density channels per the ionisation-bound LyC leakage mechanism, however in the direction of the halo-filament intersection, the I-front is bound within the halo. Therefore, in this instance a mostly confined HII region has been produced. A secondary clump within the host minihalo located in close proximity to the SF region is discernible in the first 3 rows of Fig. \ref{fig:multi_slice}. This clump is only marginally disrupted by the Pop III star but overall retains its high density throughout the stars MSL. As the Pop III star resides at the lower-end of the Pop III mass range, a weaker I-front is driven through the minihalo that is unable to revert to R-type for most of the MSL. The bound component of the HII region evolves as follows. If it is assumed that the Franco density column is static, $w = 1$ and follows the power-law $r^{-w}$ \citep{1990ApJ...349..126Franco}, then the radius of the I-front evolves to good approximation as 
\begin{equation}
    R(t) = R_s \left\{ 1 + \mathrm{W}_0 \left[-\mathrm{exp}\left(-\frac{r_c t}{R_s t_{\mathrm{rec, core}}}\right)\right]\right\}
\end{equation}
where $R_s = \mathrm{L/K}$ is the Str\"{o}mgren radius \citep{1939ApJ....89..526S} and $\mathrm{W}_0(x)$ is the principal branch of the Lambert function W. Additionally, $L \equiv \dot{N}_\gamma / (4\pi n_c r_c)$ and $K \equiv n_c r_c C_{\alpha \beta}$, where $C$ is the clumping factor. \cite{wan04} demonstrate the solution of $\mathrm{W}_0(x)$ for a bound I-front. The mass of the minihalo allows for an HII region to form in this manner, consistent with findings of \cite{wet08a}. If the minihalo mass was greater, then outward radiative pressure would not be able to compete with the downward baryonic pressure from the larger potential well and the resultant HII region would have a negligible radius (too small to be resolved). This minihalo represents an intermediate case between a trapped and unconfined HII region, where it remains confined for the majority of the MSL, breaking through the neutral shell as an R-type front only at the end of the MSL. The pressure within the HII region evacuates the central region driving the temperature past T$ = 20,000$ K whilst simultaneously reducing the density to an average of n$_{H} \sim 0.1$ cm$^{-1}$. The density reduction is enhanced at the north and south poles of the minihalo where the structure borders on the void. As the I-front transitions to a R-type front at these locations, the evolution of the I-front may reduce to a simple unbound solution 
\begin{equation}
    r_I = r_0 (1 + 2t/t_{\mathrm{rec, core}})^{1/2},
\end{equation}
where T$_{\mathrm{rec, core}}$ is the recombination time in the core if it is assumed that the the photon output per unit time, $\dot{N}_{\gamma} = 16 \pi r^3_0 n^2_0 C_{\alpha \beta} / 3$ and $r(0) = r_0$, n$_0$ is the gas number density at the characteristic radius, $r_0$ \citep{MELLEMA2006374}. Additionally, this holds only for static Franco density fields, i.e. $w = 2$, but provides a good approximation.

The MSL of the Pop III star ends after T$_{\textrm{life}}$ has elapsed, whereby the SN explosion occurs after removal of the star particle, carrying with it enriched material that propagates throughout the diffuse HII region. The energy and metal contents of the SN are deposited into a region $10$ pc in radius centred on the coordinates of the star particle.  Approximately 4.5 Myr after the SN explosion, the shock front impacts the secondary clump and violently disrupts it, reducing its density. The clump falls into the recollapsing region of the SN remnant after a period of brief recovery $\sim 54$ Myr after the explosion. The SN ejecta exhibits distinct bipolar outflow from the central region, as it is forced through regions of lower density characterised by HII breakout that is observed at the north and south poles of Fig. \ref{fig:multi_slice}. Finally after $190$ Myr, the halo begins to recover and initiate the process of recollapse. The selected host minihalo represents a unique scenario where the binding energy of the halo and the energy of the SNe are finely balanced. The minihalo of mass M$_{\mathrm{halo}} = 4.8 \times 10^{5}$ M$_{\odot}$ at $z = 19.10$ yields a gravitational binding energy 
\begin{equation}
    E_\mathrm{{bind}} = -\frac{3}{5}\frac{G \mathrm{M}_{\mathrm{halo}}^2}{R_{\mathrm{vir}}} = 7.45 \times 10^{50}, \mathrm{ergs}
\end{equation}
where $G$ is the gravitational constant and R$_{\mathrm{vir}}$ is the virial radius given as approximately $145$ pc. The SNe energy $E_{\mathrm{SN}} = 5 \times 10^{50}$ ergs. Therefore in this instance $E_\mathrm{{bind}} \lesssim E_{\mathrm{SN}}$, although they can be approximated as the same with regards to the method of enrichment. In theory, this presents an opportunity for both EE and/or internal enrichment (IE) being realised, or at the very least increasing the timescale for recollapse to occur significantly, when compared to the same configuration with a higher mass halo ($E_\mathrm{{bind}} \gg E_{\mathrm{SN}}$) or a more energetic SNe ($E_\mathrm{{bind}} \ll E_{\mathrm{SN}}$). The first case would be a strong candidate for complete IE, whereas the second for complete EE rather than a mixture of the two. 

\subsection{Long-term Remnant Evolution}

\begin{figure*}
    \centering
    \includegraphics[scale=0.5]{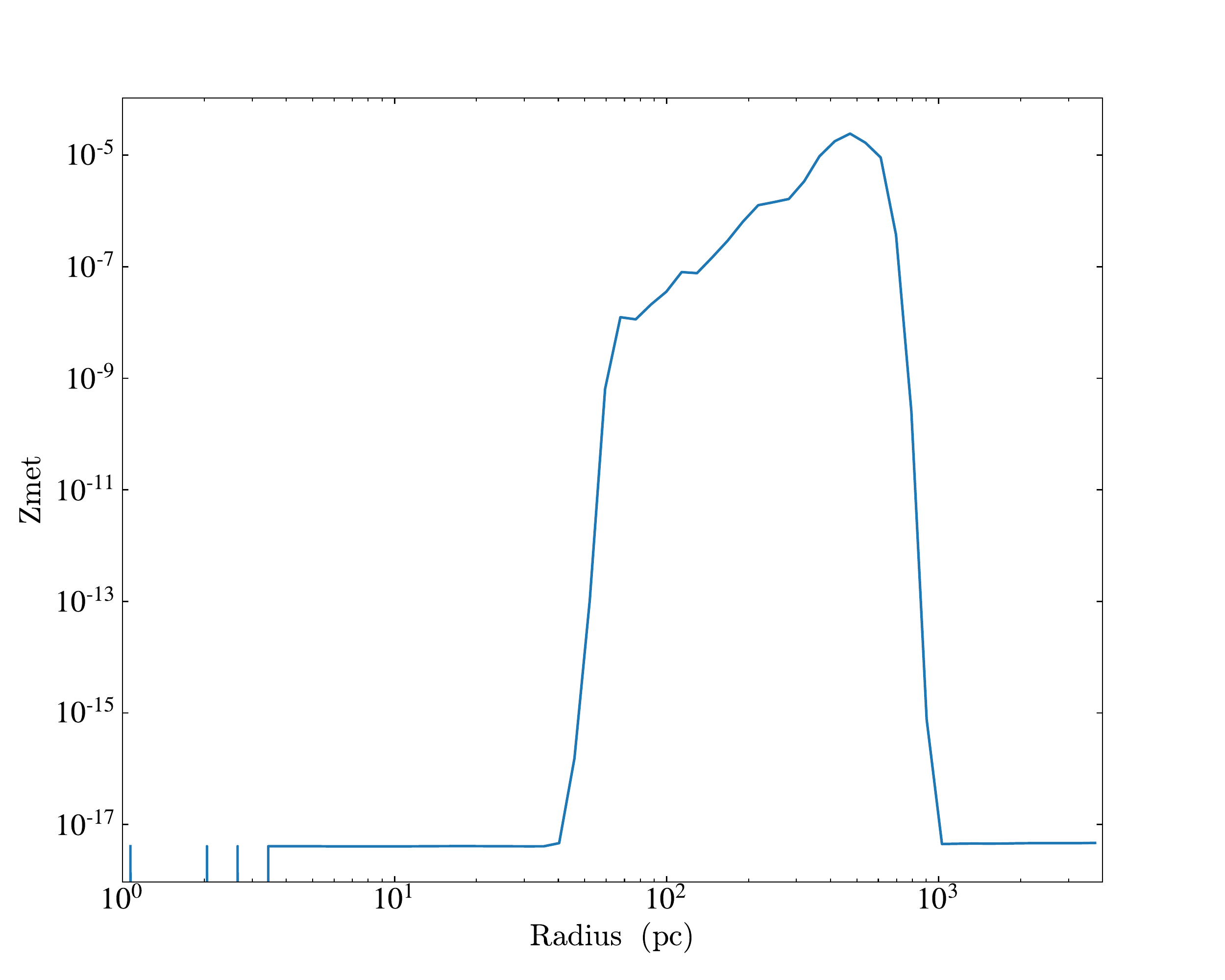}
    \includegraphics[scale=0.5]{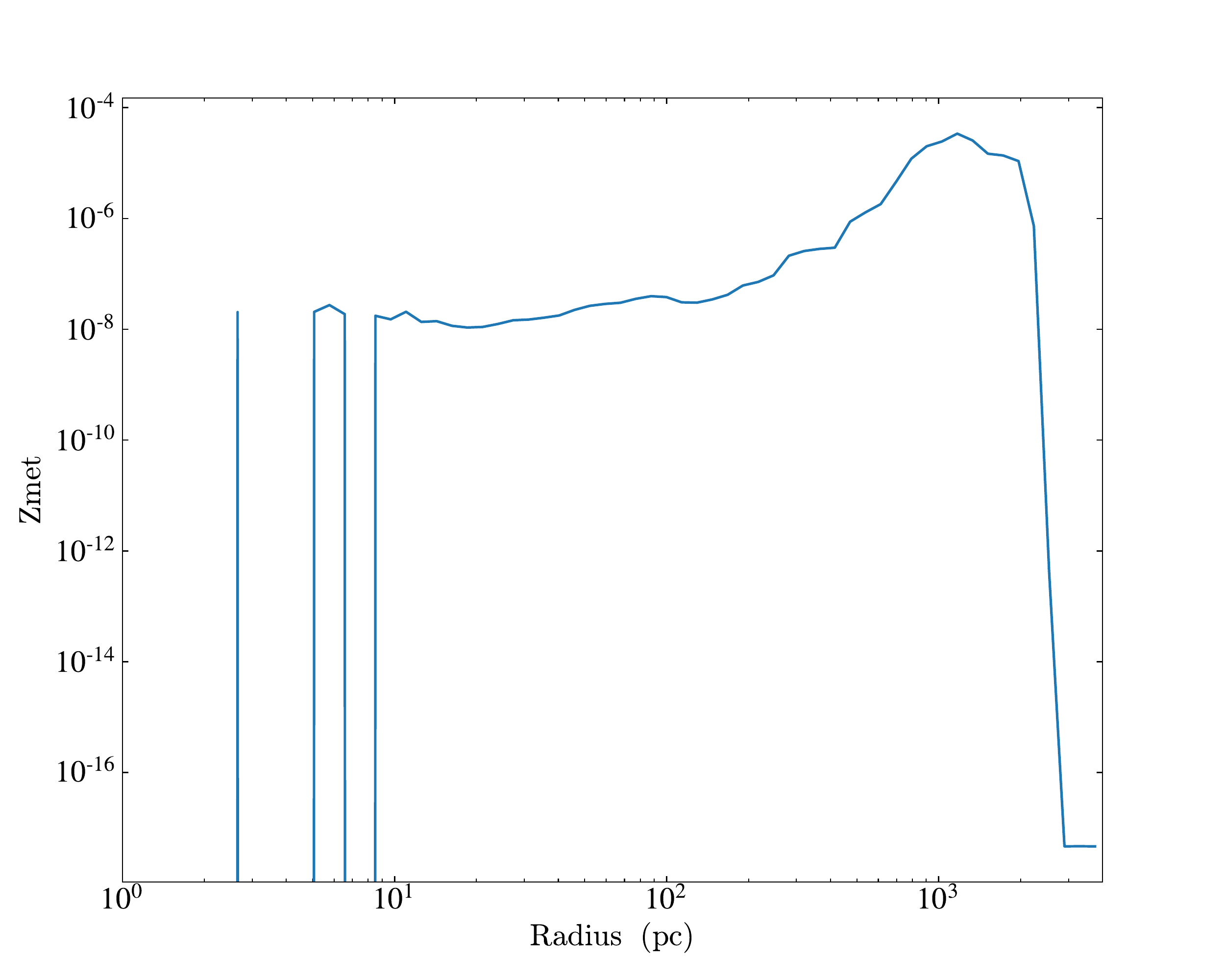}
    \label{fig:radial_met}
    \caption{Spherically averaged radial profile of the SN remnant in both simulations 112 Myr post-SNe displaying the metallicity weighted by cell mass as a function of radius, centred on the coordinates of the SNe. Top: Main faint-CCSN simulation. Bottom: Normal CCSN simulation. The increase of metallicity as the radius grows demonstrates the bulk of the metals being transported out of the central region. Note the CCSN plot that demonstrates a more consistent distribution of metals due to the greater metal ejecta mass.}
\end{figure*}

The unconfined nature of the HII region gives rise to the visually distinct bipolar outflow of metal, forced through the poles where the minihalo borders the void and extending further than 2 kpc/h after 112 Myr. The consequences of such a blowout are extremely significant for the subsequent chemical enrichment of the remnant. The metallicity within the central 100 pc region is radially reduced almost uniformly from $Z = 10^{-15}\ Z_{\odot}$ right at the coordinates of the dead star, as the over dense region lying in close proximity falls into this region. In fact, the bulk of the metal ejecta is transported out of the minihalo within the bipolar outflow into the void. Evident in Fig. \ref{fig:radial_met}, the highest average metallicity at this time is located $\sim 600$ pc away in the extrema of the ejecta, and reaches $Z > 10^{-6}\ Z_{\odot}$. Although the metallicity is relatively high within these regions, the density is so low that the primary reactions governing water formation and other heavy molecule formation are effectively stifled. For any significant quantities of water to form, a recollapsing region exceeding n$_\mathrm{H} = 10^{3}$ cm$^{-3}$ and enriched to a minimum metallicity ($Z \sim 5\times 10^{-5} Z_{\odot}$ for a fiducial value) is required to initiate runaway recollapse. 

\begin{figure*}
    \centering
    \includegraphics[width=1.0\linewidth]{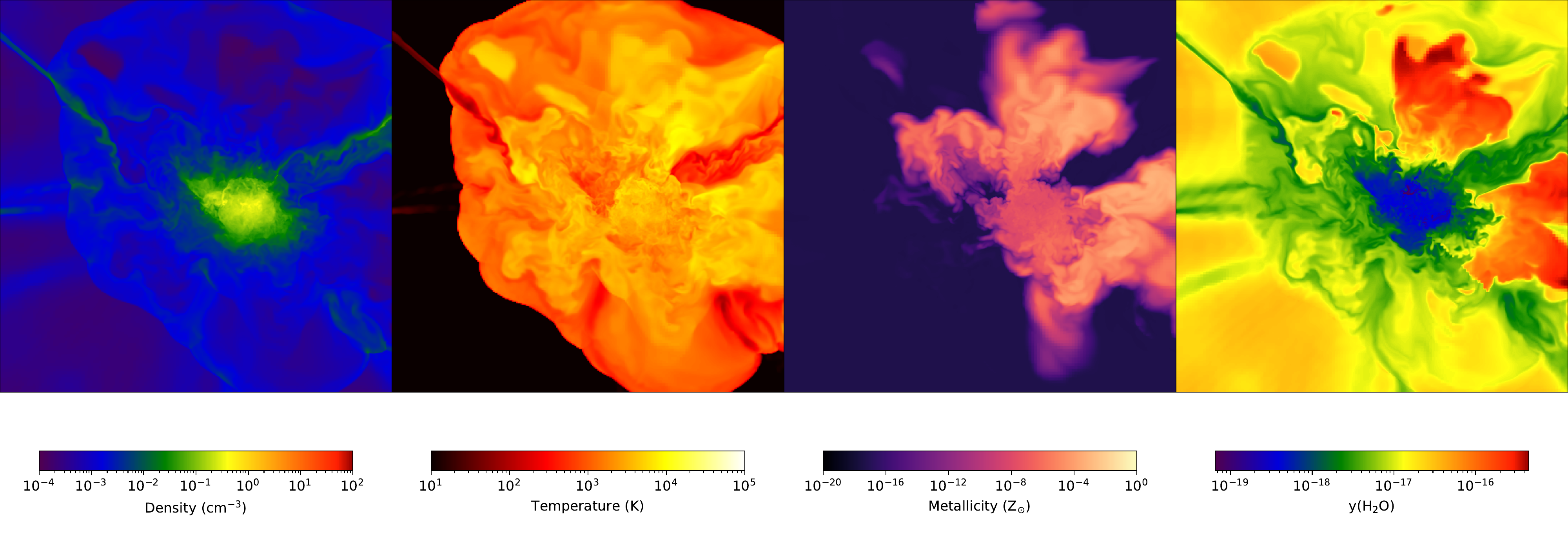}
    \includegraphics[width=1.0\linewidth]{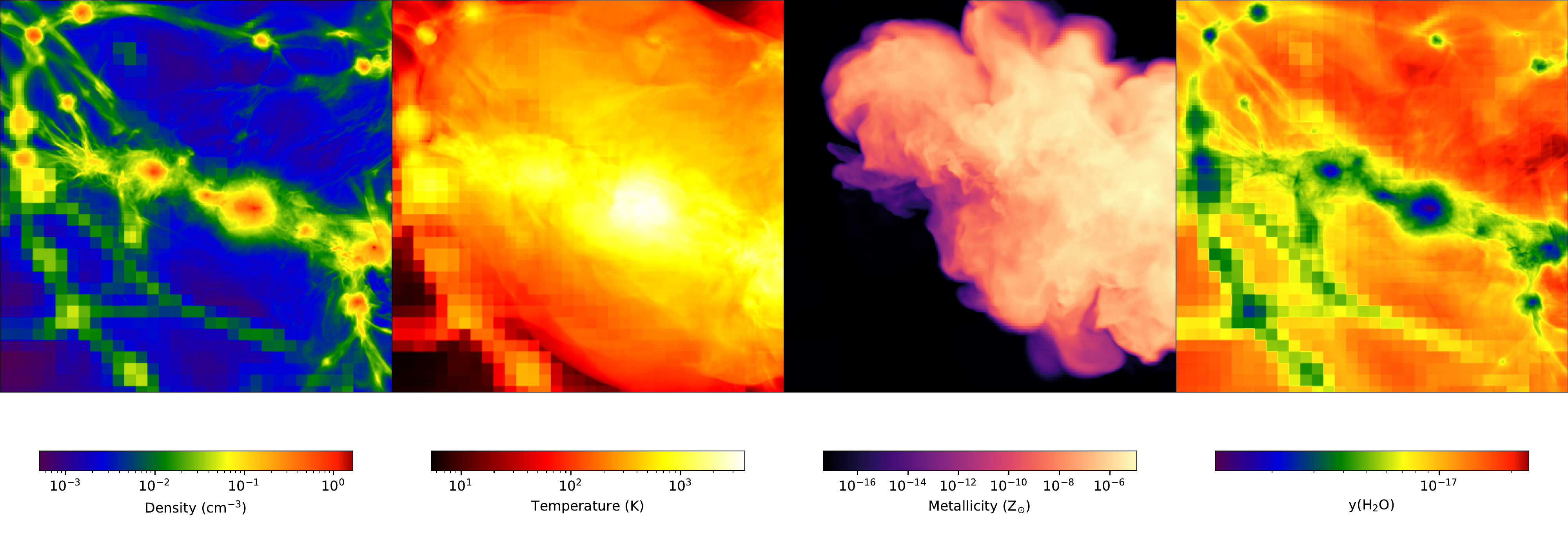}
    \caption{Top: Slices of density, temperature, metallicity, and water abundance centred on the coordinates of SF 192 Myr post-SNe. Bottom: Density weighted projections centred on the same coordinates at the same epoch. Each figure has a side length of 4 kpc comoving.}
    \label{fig:out_389_slice_proj}
\end{figure*}

Immediately prior to the recollapse of multiple small regions throughout the remnant, the distribution of metals within this region are visible in Fig. \ref{fig:out_389_slice_proj}. Perhaps the most obvious feature within the top series of slice plots is the distribution of higher water abundance (although marginal) at the same position as the relatively higher metallicities. Specifically, the darker orange/red of water abundance matches the position of the brighter spots in the adjacent metallicity plots, whereas the centre has effectively 0 water abundance denoted by the dark blue, and is in the same location as the darker region of metallicity. The region of high density that dominates the top left plot is the consequence of the high density region that lay in close proximity to the blast, falling into the central region 4.5 Myr after the explosion. 192 Myr post SNe, the clump reaches a faux equilibrium with the surrounding remnant, remaining in essentially the same state from fall-in until the onset of recollapse. The series of projection plots in the lower half of Fig. \ref{fig:out_389_slice_proj} gives a greater representation of the distribution of the dense knots that permeate the remnant. Forming a filamentary structure with the aforementioned clump at the centre extends horizontally across the figure, the collapsing regions remain at low metallicity. The incorporation of metals into the surrounding pristine minihalo structure via mixing has been studied by \cite{cen2008lower}, and in some instances 90 \% of high density gas inside the $10^6\ \mathrm{M}_{\odot}$ minihalo is enriched to only 3 \% of the surrounding metallicity by $z = 6$ suggesting that metal-free Pop III SF may be possible down to lower redshifts. In our minihalo, the mass at SF is an order of magnitude lower and therefore the apparent inability of metal-enriched material to mix with pristine gas may be amplified, as evident in Fig. \ref{fig:out_389_slice_proj}.

\subsection{Water Formation in a Low-Mass Minihalo}

In our simulation, there is negligible water formation in the remnant of the SN after significant time (192 Myr) has passed. We define a significant amount of time as the dynamical time when recollapse would be expected in a halo of mass M $\sim 1\times 10^6\ \mathrm{M}_{\odot}$ which is approximately 80 Myr. There are regions within the remnant region that have an elevated water abundance with respect to the background value by a few orders of magnitude, indicating that the remnant is not completely devoid of water. These values however, are at a maximum $y\_\mathrm{H}_2\mathrm{O} \sim 10^{15}$ and exceedingly low. The metallicity in these "high" water abundance regions has a minimum value of Z$\ = 10^{-9}\ \mathrm{Z}_{\odot}$, below this metallicity value the water abundance does not increase further than its background initialisation values. Furthermore, the high metallicity regions exist at significant distance from the central overdense region. This suggests that a significant fraction of metal would need to halt further outward expansion and fall back onto the core before the potential for mixing could occur, let alone effective mixing of metal with primordial gas. 

A simulation was initialised with identical initial conditions to ensure the characteristics of the minihalo were unchanged. An identical mass ($13$ M$_{\odot}$) Pop III star was inserted at the same time described above, only in this instance the SN model was altered to represent a `nominal' CCSN, i.e. the energy was E$_{\mathrm{SN}} = 10^{51}$ ergs and the SN model outputs almost 10 times the metals at M$_{\mathrm{met}} = 0.784\ \mathrm{M}_{\odot}$ as the faint-SNe. The progenitor star evolves in an identical fashion and so the initial conditions for the explosion to occur in are consistent between the two simulations. After the SNe occurs, the dense clump lying in close proximity is impacted by the shock wave and is partially disrupted. The clump falls into the central region in a similar fashion after a period of brief recovery. After 120 Myr, the same filamentary structure that runs horizontally connecting regions of high density becomes visible, with the pristine clump of gas forming the centre. 

\begin{figure*}
    \centering
    \includegraphics[width=1.0\linewidth]{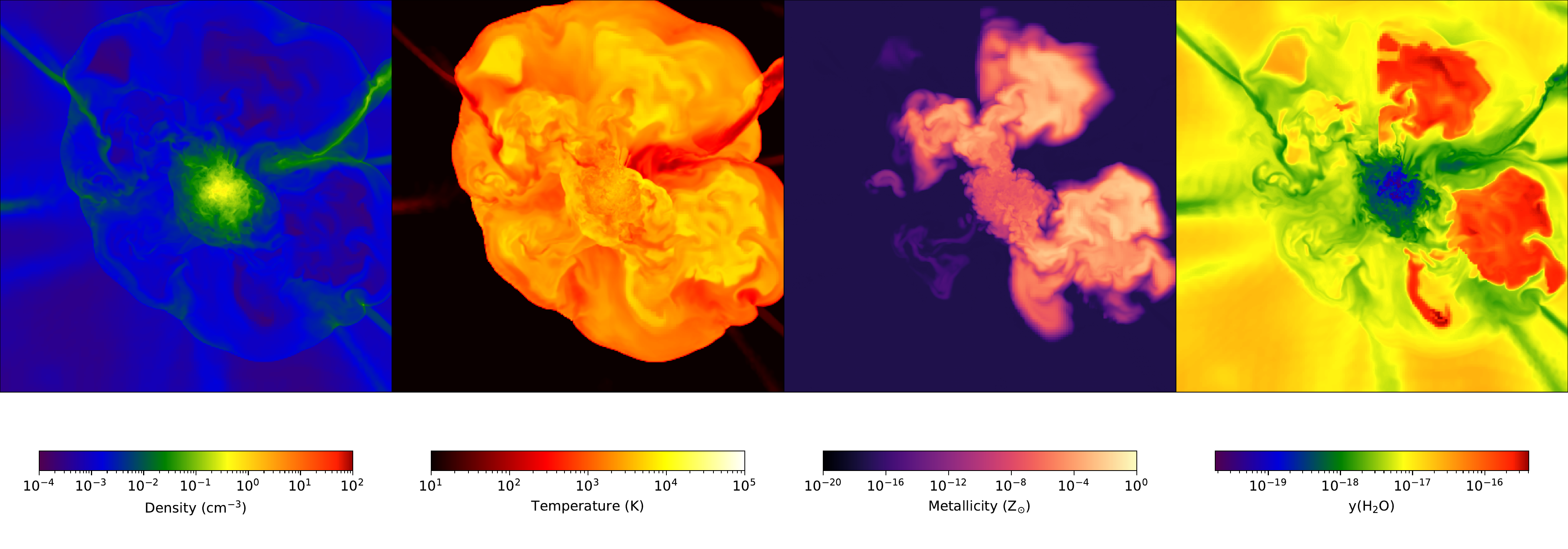}
    \includegraphics[width=1.0\linewidth]{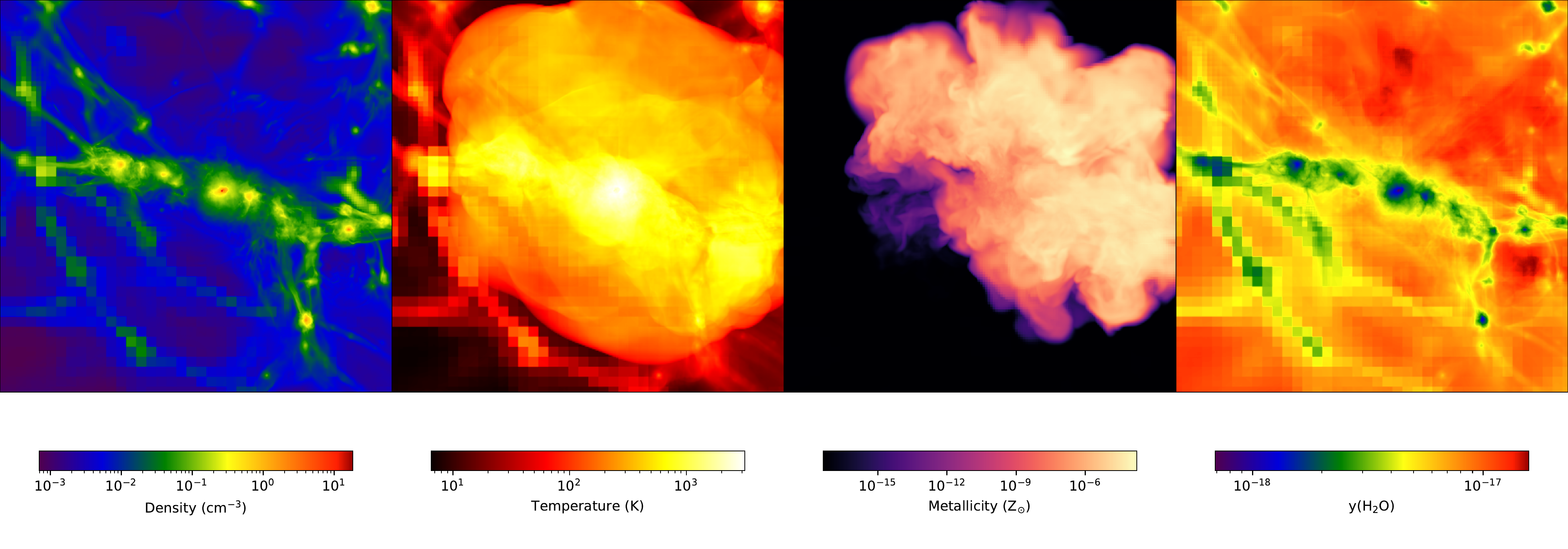}
    \caption{Similar to Fig. \ref{fig:out_389_slice_proj}, but for a normal CCSN. Top: Slices of density, temperature, metallicity, and water abundance centred on the coordinates of SF 120 Myr post-SNe. Bottom: Density weighted projections centred on the same coordinates at the same epoch. Each figure has a side length of 4 kpc comoving.}
    \label{fig:CCSN_norm_slice_proj}
\end{figure*}

Fig. \ref{fig:CCSN_norm_slice_proj} displays the state of the remnant and clearly shows this structure forming. Although taken at a different epoch, comparing with Fig. \ref{fig:out_389_slice_proj} the similarities are obvious. The bulk of the metals are transported away from the central region, and marginal water formation occurs within the outer extremities of the remnant where the densities are the lowest. In addition, the collapsing dense substructures that make up the horizontal filament have the lowest water abundance values throughout the remnant at y\_H$_2$O $< 10^{-18}$, over two orders of magnitude less than the lower density regions.

\begin{figure*}
    \centering
    \includegraphics[scale=0.3]{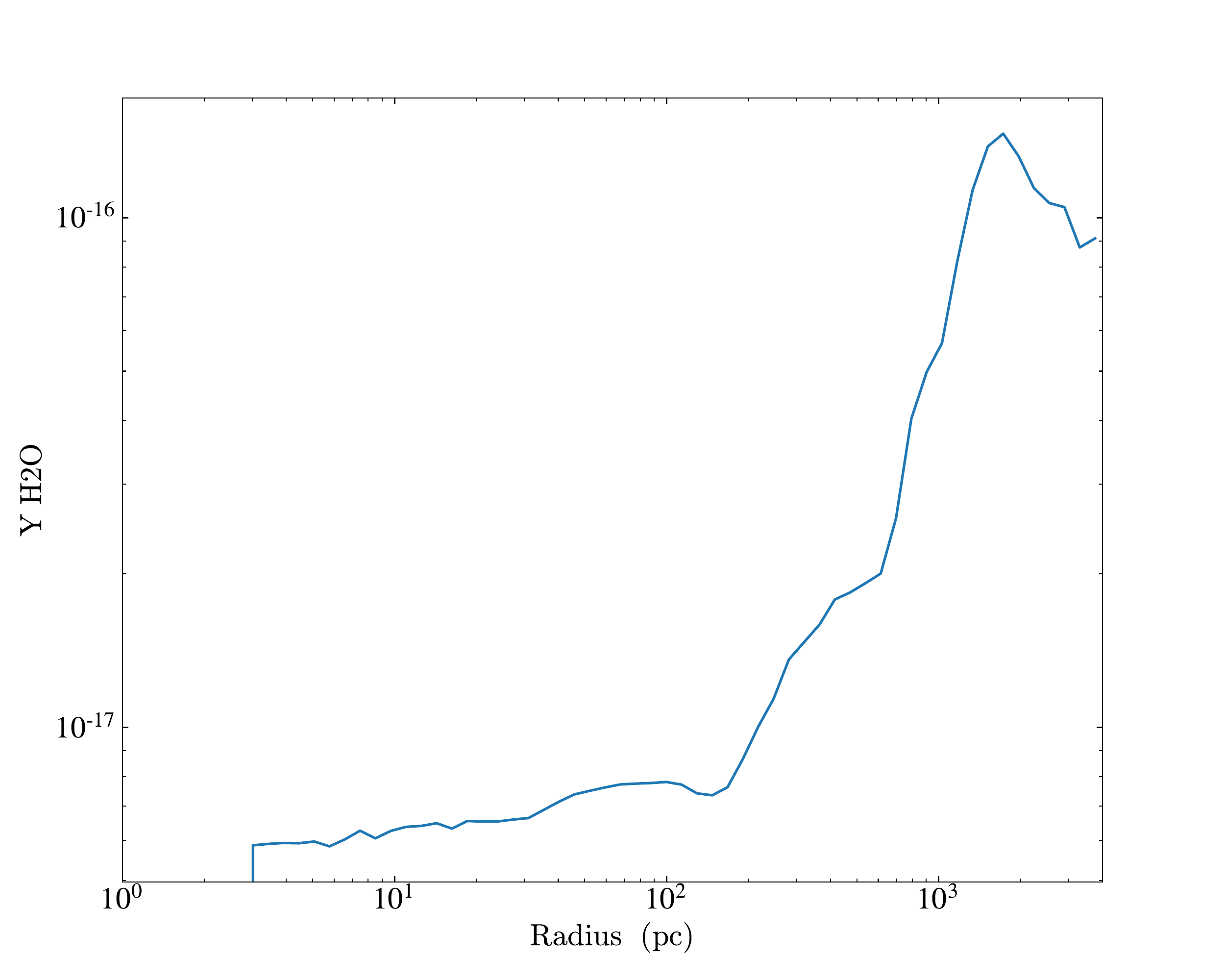}
    \includegraphics[scale=0.3]{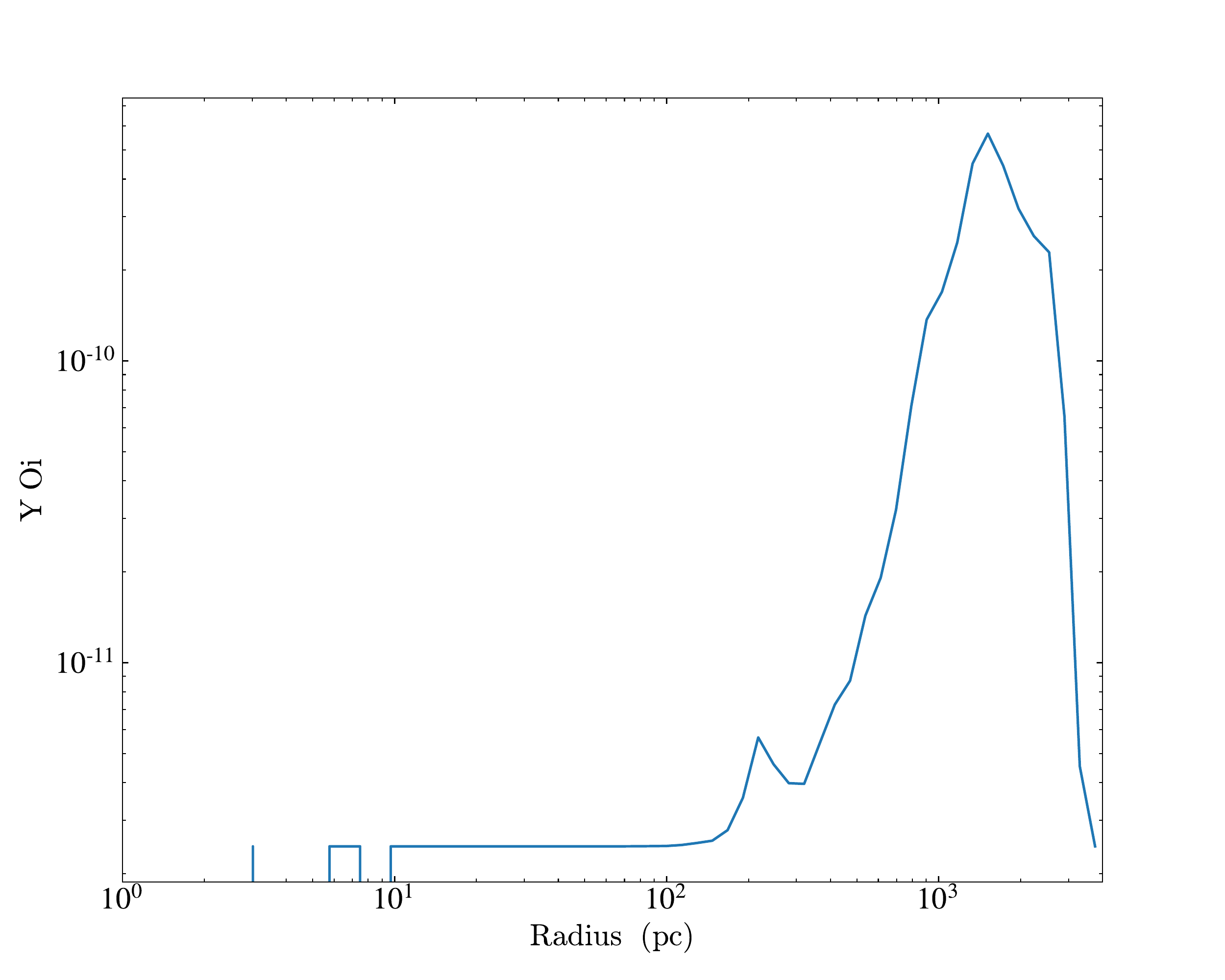}
    \includegraphics[scale=0.3]{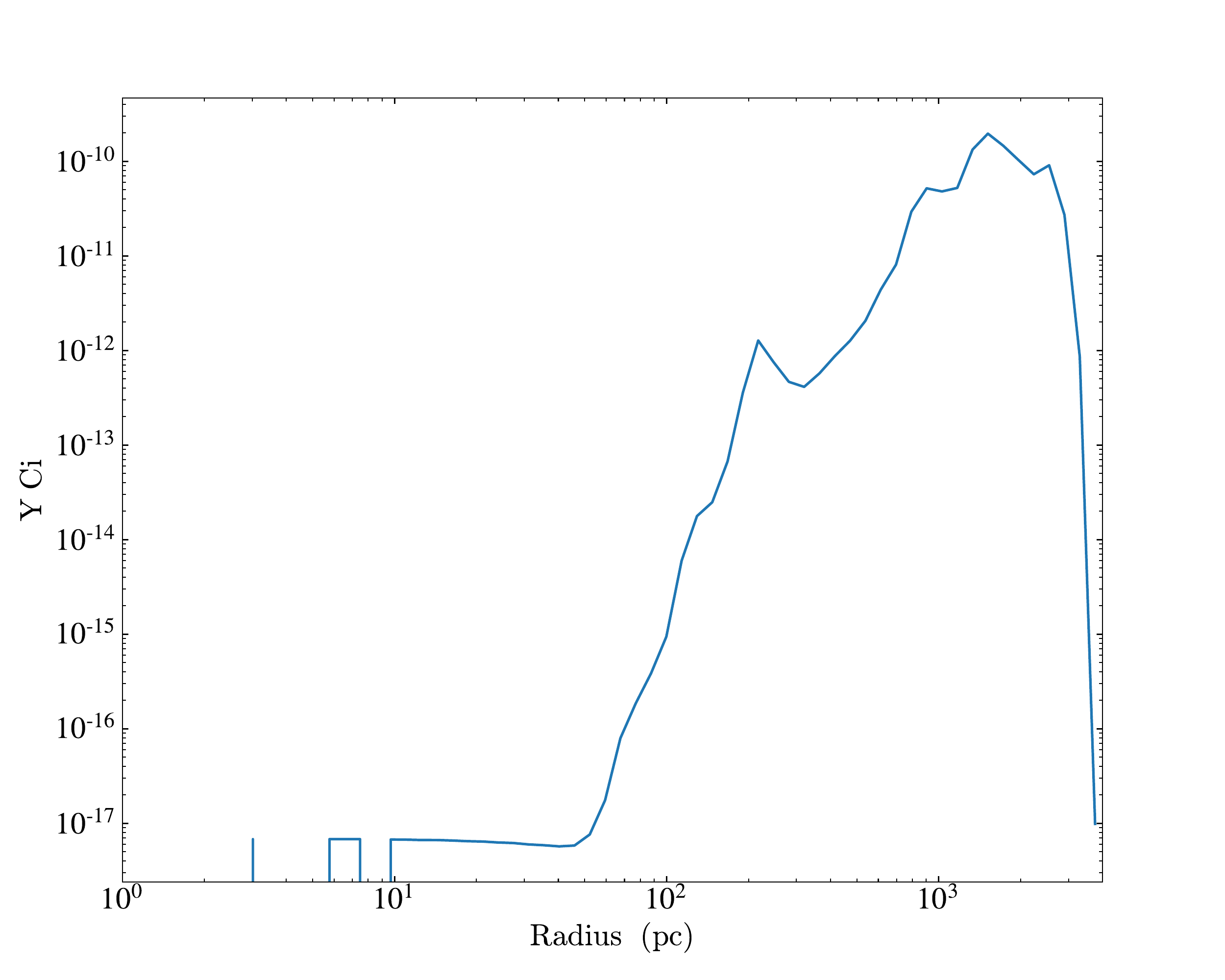}
    \includegraphics[scale=0.3]{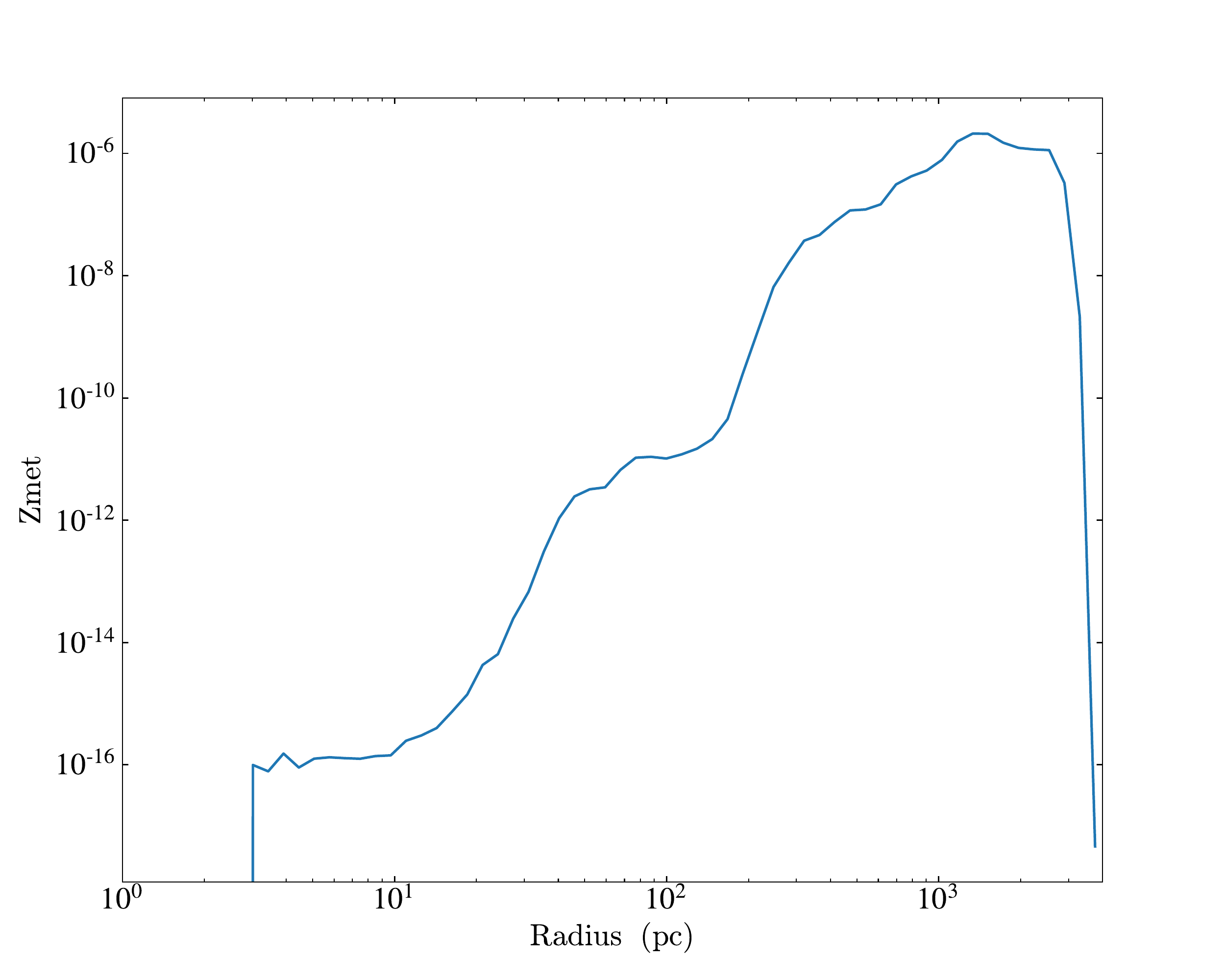}
    \caption{Spherically averages radial profile of the SN remnant 192 Myr post-SNe of the major species and metallicity.}
    \label{fig:192_radial_plots}
\end{figure*}

\section{Discussion and Conclusions}

Here we present the first results in a series where we attempt to find the ideal conditions for water formation in the remnant of Pop III supernovae. For the first time, we solve all the relevant metal chemistry that governs complex molecule formation and include a detailed prescription for dust chemistry, in addition to mass dependant Pop III SN yields in a self-consistent manner. In the target minihalo ($10^{5}$ M$_{\odot}$) initialised from cosmological initial conditions, pristine gas collapses via H$_2$ cooling and forms a 13 M$_{\odot}$ Pop III star at $z = 19.02$. We simulate the MSL of the star using the  \textsc{MORAY} radiative transfer algorithm. The Pop III star is modelled to have significant fallback onto the core at the point of explosion, reducing the energy of the explosion to $5 \times 10^{50}$ ergs and the metal ejecta to just 0.097 M$_{\odot}$. The CCSN exhibits striking bipolar outflow from the central region and forces the bulk of the metal ejecta into the void. The simulation concludes when the dense substructures throughout the remnant begin catastrophic recollapse. A summary of the primary findings from this simulation are as follows:
\begin{enumerate}
    \item The low-mass of the minihalo allows for the HII region generated by the Pop III star to escape the virial radius near the end of the stars lifetime, transitioning from a D-type to R-type in the direction of the voids. 
    \item The star explodes as a CCSN and the metal ejecta is forced through the regions where LyC leakage occurs, whilst remaining confined where the minihalo primary structure meets the bordering filaments. This creates a bipolar outflow of metals and efficiently transports the metal ejecta out of the minihalo. 
    \item A dense, pristine clump of gas lies close to the SN and is impacted 4.5 Myr later. This clump is only partially disrupted by the shock and recovers quickly. The clump falls into the central region $\sim$ 54 Myr later.
    \item The metallicity of the core reaches a minimum value of $10^{-14}$ Z$_{\odot}$ which increases to a maximum value of $10^{-6}$ Z$_{\odot}$ approximately 900 pc from the core. This suggests firstly that the majority of metals are transported away from the core in the explosion and secondly, that the mixing of metals into the primordial material that the infalling clump provides is very inefficient. 
    \item 120 Myr post-SNe, the formation of a horizontal filament permeated by areas of high density with the collapsing infalling clump lying at the centre becomes distinctly visible. These areas have a lower metallicity and also have a few orders of magnitude lower water abundance than that of the surrounding higher metallicity regions that exist at lower density. This again suggests that mixing is extremely inefficient. 
    \item Water formation occurs within low density regions at the outskirts of the remnant, although the abundance peaks no higher than y\_H$_2$O $= 10^{-15}$ which is almost negligible in the context of second generation star and protoplanetary disk formation. The pristine clump lying at the centre of the remnant loiters in a faux equilibrium, not incorporating metals into itself and actually suppressing complex chemical reactions.  
    \item An identical simulation up until SN was initialised where the energy was doubled and the metal ejecta increased by a factor of 10. The chemo-thermal evolution of the remnant was very similar, with the majority of metal ejecta being forced into the void and the dense clump falling into the central region.
\end{enumerate}
The configuration of this minihalo mass containing within it two overdense regions at SF time conspire to create a unique situation where metal mixing in the centre of the remnant is inefficient, and therefore complex chemical interactions are suppressed. As soon as the HII region of a Pop III star is able to overcome the minihalo, any SNe has the ability to blow the majority of its metal ejecta out into the void. In this instance, we chose to model a faint-CCSN to give the minihalo the greatest chance of recovery within a Hubble time as this represents a scenario with the least energy output. Even so, the halo is blown apart and it takes at least 190 Myr for some semblance of recollapse to occur. If a minihalo had lay in close proximity to the host minihalo, EE may be successful in creating the conditions for water formation \citep{smith2015first}. As this was not the case here, IE was the only avenue for water formation to occur, however this never happened as the pristine clump fell onto the central region of the remnant and diluted the metallicity. Since the majority of metals were ejected past the virial radius, the presence of the clump may only have enhanced the water formation rate (or lack thereof) that we see in the core. Even when increasing the metal ejecta mass by a factor of 10 from a normal CCSN, the explosion energy is doubled and the result is the same (Fig. \ref{fig:CCSN_norm_slice_proj}). The minihalo is blown out and the majority of metals are transported into the void. For IE to be viable and recollapse to occur on short timescales, the HII region must be confined to the virial radius of the minihalo. This would trap the SNe, which would reach equilibrium and initiate recollapse earlier whilst retaining the bulk of the metal ejecta. \cite{10.1093/mnras/sty2984} demonstrate that a minihalo with mass $1.77 \times 10^{6}$ M$_{\odot}$ is able to confine a $13$ M$_{\odot}$ Pop III stars HII region, suggesting that a lower limit can be placed on a minihalo that is able to self-enrich itself via IE as $\sim 10^{6}$ M$_{\odot}$. 
The dynamics of each minihalo are unique, and therefore this may not be so much a set rule for all minihaloes at the lower end of the mass spectrum as a rule for this specific configuration. What is clear however, is that as the minihalo mass increases so does the likelihood that IE dominates and the chances of water formation also increases. Future studies are required on the matter, and will be able to probe what happens when the configuration changes slightly. Increasing the mass of the minihalo at SF to the upper half of $10^{5}$ M$_{\odot}$, or alternatively increasing the mass of the Pop III progenitor to inject more metals into the surrounding area may prove to have profound effects on the end state of the remnant and the ability of the recollapsing clumps to mix metals into themselves. Finding the correct balance between SN explosion energy, metal ejecta mass, and the host minihalo mass may be the key to discovering the configuration which promotes the formation of a water abundant protoplanetary disk around a second-generation star, and therefore the emergence of the first wet rocky planets.

\section*{Acknowledgements}

C. T. D. Jessop was supported by STFC grant 18379. All numerical simulations were performed on the Sciama HPC cluster, supported by the University of Portsmouth and the Institute of Cosmology and Gravitation. Both computation and analysis of the simulations in the above text were carried out with the \textsc{enzo} and  \textsc{yt} codes, both of which are publicly available and maintained by a collaboration of scientists internationally. 

\section*{Data Availability}

The data used for this article will be shared upon reasonable request to the author.



\bibliographystyle{mnras}
\bibliography{refs.bib} 








\bsp	
\label{lastpage}
\end{document}